\documentstyle[eqsecnum,pre,aps,epsf,epsfig]{revtex}

\begin{document}
\draft


\title{Effects of surfaces on resistor percolation
}
\author{Olaf Stenull, Hans-Karl Janssen and Klaus Oerding\footnote{Present Address: SAP AG, Neurottstra{\ss}e 16, 69189 Walldorf, Germany} 
}
\address{
Institut f\"{u}r Theoretische Physik 
III\\Heinrich-Heine-Universit\"{a}t\\Universit\"{a}tsstra{\ss}e 1\\40225 
D\"{u}sseldorf, Germany
}
\date{\today}
\maketitle

\begin{abstract}
We study the effects of surfaces on resistor percolation at the instance of a semi-infinite geometry. Particularly we are interested in the average resistance between two connected ports located on the surface. Based on general grounds as symmetries and relevance we introduce a field theoretic Hamiltonian for semi-infinite random resistor networks. We show that the surface contributes to the average resistance only in terms of corrections to scaling. These corrections are governed by surface resistance exponents. We carry out renormalization group improved perturbation calculations for the special and the ordinary transition. We calculate the surface resistance exponents $\phi_{\mathcal S \mathnormal}$ and $\phi_{\mathcal S \mathnormal}^\infty$ for the special and the ordinary transition, respectively, to one-loop order.
\end{abstract}
\pacs{PACS numbers: 64.60.Ak, 64.60.Fr, 68.35.Rh, 05.70.Np}


\section{Introduction}
Percolation~\cite{bunde_havlin_91_stauffer_aharony_92} is the perhaps simplest model for the irregular geometry which occurs in disordered media. Percolation is intuitively appealing and it has a large variety of applications. Moreover, percolation is the prototype of a geometric phase transition. The continuing interest in percolation results in an abundance of publications year after year. In particular the purely geometric aspects of percolation have been studied extensively. From the current perspective, non equilibrium properties of percolation, like transport on percolation clusters, are of growing interest. Of course these issues are typically more challenging than the equilibrium properties.

Random resistor networks (RRN) play a major role in the study of transport on percolation clusters. A RRN is simply a bond percolation model in which the occupied bonds are assigned a finite, nonzero conductivity. Commonly studied in the theory of RRN is the average resistance $M_R (x, x^\prime)$ between two connected ports $x$ and $x^\prime$ when an external current is inserted at $x$ and withdrawn at $x^\prime$. It was found~\cite{harris_fisch_77,dasgupta_harris_lubensky_78} that $M_R (x, x^\prime)$ scales at the percolation point as
\begin{eqnarray}
M_R (x, x^\prime) \sim \left| x - x^\prime \right|^{\phi/\nu} \ ,
\end{eqnarray}
with $\phi$ being referred to as resistance exponent and where $\nu$ is the correlation length exponent of the percolation universality class. 

The renormalization group provides a powerful and elaborate framework to investigate RRN analytically. In particular a field theoretic approach based on the seminal work of Stephen\cite{stephen_78} and Harris and Lubensky~\cite{harris_lubensky_87b} has prooved to be fruitful. Using this approach the resistance exponent has been calculated~\cite{stenull_janssen_oerding_99} to second order in $\epsilon = 6 - d$, where $d$ is the spatial dimension. Moreover, the approach has been used to compute $\phi$ for continuum percolating networks~\cite{lubensky_tremblay_86}, $\phi$ for diluted networks of nonlinear resistors~\cite{harris_87,janssen_stenull_oerding_99,janssen_stenull_99}, several fractal dimensions characterizing percolation clusters~\cite{harris_87,janssen_stenull_oerding_99,janssen_stenull_99}, an entire family of multifractal exponents for the moments of the current distribution in RRN~\cite{park_harris_lubensky_87,stenull_janssen_2000a,stenull_janssen_2000b}, etc.

Though a field theory of boundary critical phenomena has been established (for background on the field theoretic approach to boundary critical behaviour see Refs.~\cite{diehl_86,diehl_97}) and successfully applied to geometric percolation~\cite{JSS_88,diehl_lam_89} the field theoretic approach to RRN has not yet been extended to include surfaces. In this paper we present such an extension based on the approach of Stephen and Harris and Lubensky. We consider the, at least from the standpoint of field theory, simplest geometry which comprises a surface, viz.\ a semi-infinte geometry. The central question addressed by this paper is: What is the critical behavior of $M_R (x, x^\prime)$ when the ports $x$ and $x^\prime$ are located on the surface and how does the surface contribute to it?

The plan of presentation is the following: the paper has two main sections, Secs.~\ref{model} and  \ref{rga}. In Sec.~\ref{model} we provide some background on the phenomenology of resistor percolation. Then we develop a field theoretic model which has its manifestation in a Landau-Ginzburg-Wilson type Hamiltonian. We show that the term in this Hamiltonian steaming from the conductance of the surface bonds is irrelevant in the sense of the renormalization group. In Sec.~\ref{rga} we present the core of our renormalization group analysis. We calculate correction to scaling exponents for the average resistance which are associated with the irrelevant surface term. These are then compared to the Wegner exponent for percolation. Section~\ref{Conclusions} contains our conclusions. Details of the calculations are relegated to three appendices.

\section{The Model}
\label{model}

\subsection{Semi-infinite random resistor networks}
Consider bond percolation on a semi-infinite lattice in $d$ dimensions bounded by a $(d-1)$-dimensional plane. A lattice site $i$ is either located off the surface, i.e., $i \in {\mathcal B \mathnormal} = \left\{ x = \left( x_{\|} , z \right) \left| \   x_{\|} \in {\rm{\bf Z}}^{d-1}, z \in {\rm{\bf N}}/\left\{ 0 \right\} \right\} \right.$, or on the surface, i.e., $i \in {\mathcal S \mathnormal} = \left\{ x = \left(  x_{\|}, 0 \right) \left| \ x_{\|} \in {\rm{\bf Z}}^{d-1} \right\} \right.$. Each bond between nearest neighbors on the surface is occupied by a resistor of conductance $\sigma_{\mathcal S \mathnormal}$ with probability $p_{\mathcal S \mathnormal}$ or unoccupied with probability $1-p_{\mathcal S \mathnormal}$. The bonds between all other nearest neighbor pairs are occupied by resistors of conductance $\sigma$ with probability $p$ or empty with probability $1-p$.

\begin{figure}[h]
\epsfxsize=8.4cm
\centerline{\epsffile{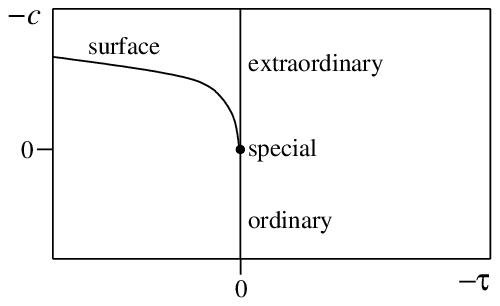}}
\caption[]{\label{phaseDiagram}Schematic phase diagram for semi-infinite percolation. The horizontal axis corresponds to $-\tau \sim p - p_c$. The vertical axis corresponds to c, whose negative is a measure of the surface enhancement. The lines labeled {\sl ordinary}, {\sl extraordinary} and {\sl surface} indicate continuous phase transitions that have been given these names by Lubensky and Rubin\cite{lubensky_rubin_75}. The lines meet at a tricritical point that represents the {\sl special} transition.}
\end{figure}

Different phases can be distinguished depending on the values of $p$ and $p_{\mathcal S \mathnormal}$\cite{diehl_lam_89}. A sketch of the phase diagram is given in Fig.~\ref{phaseDiagram}. The parameter $\tau \sim p_c - p$ indicates if $p$ exceeds its critical value $p_c$. $p_{\mathcal S \mathnormal}$ may be enhanced with respect to $p$. The parameter $c \sim p_{\mathcal S \mathnormal ,c} \left( p = p_c \right) - p_{\mathcal S \mathnormal}$ serves as a measure of the enhancement. Let $P_{\mbox{\scriptsize{perc}}} \left( x \right)$ denote the percolation probability that site $x$ belongs to an infinite cluster. Assume $c>0$. Crossing over from $\tau >0$ to $\tau <0$ leads to the formation of an infinite cluster. However, $P_{\mbox{\scriptsize{perc}}} \left( x \in {\mathcal S \mathnormal} \right) \leq P_{\mbox{\scriptsize{perc}}} \left( x \in {\mathcal B \mathnormal} \right)$ by virtue of the missing neighbors at the surface. The phase transition that takes place at $\tau =0$ is called {\sl ordinary} transition. Now suppose that $\tau$ is subcritical and that the system crosses over from $p_{{\mathcal S \mathnormal}} < p_{{\mathcal S \mathnormal} ,c}$ to $p_{\mathcal S \mathnormal} > p_{{\mathcal S \mathnormal} ,c}$ at fixed $p$. At this so-called {\sl surface} transition an infinite cluster forms in the vicinity of the surface but $P_{\mbox{\scriptsize{perc}}} \left( \left( x_{\|}, z \right) \right)$ falls off exponentially for $z \to \infty$. Upon increasing $p$ at fixed $p_{\mathcal S \mathnormal}$, the {\sl surface} transition is followed by the {\sl extraordinary} transition at which the exponential decay ceases to exist. The point $\left( c=0, \tau = 0 \right)$ defines a tricritical point describing the so-called {\sl special} transition.

Suppose a current $I$ is injected into a cluster at site $x$ and withdrawn at site 
$x^\prime$. The current carrying bonds constitute, apart from Wheatstone bridge type configurations, the backbone between $x$ and $x^\prime$. The power dissipated on the backbone is by definition
\begin{eqnarray}
\label{powerDef}
P=I \left( V_x - V_{x^\prime} \right) \ ,
\end{eqnarray}
where $V_x$ is the potential at site $x$. Using Ohm's law, 
\begin{eqnarray}
\sigma_{i,j} \left( V_j - V_i \right) = I_{i,j} \ , 
\end{eqnarray}
where $I_{i,j}$ is the current flowing through the bond from $j$ to $i$, it may be expressed entirely in terms of voltages as
\begin{eqnarray}
\label{powerInTermsOfV}
P=  R (x ,x^\prime)^{-1} \left( V_x - V_{x^\prime} \right)^2 = \sum_{\langle i,j \rangle} \sigma_{i,j} \left( V_i - V_j \right)^2 =  P \left( \left\{ V \right\} \right) \ .
\end{eqnarray}
Here $R (x ,x^\prime)$ is the total resistance of the backbone, the sum is taken over all nearest neighbor pairs on the cluster and $\left\{ V \right\}$ denotes the corresponding set of voltages. As a consequence of the variation principle
\begin{eqnarray}
\label{variationPrinciple1}
\frac{\partial}{\partial V_i} \left[ \frac{1}{2} P \left( \left\{ V \right\} 
\right) - \sum_j I_j V_j \right] = 0 \ ,
\end{eqnarray}
one obtains Kirchhoff's law
\begin{eqnarray}
\label{cirquitEquations}
\sum_{\langle j \rangle} \sigma_{i,j} \left( V_i - V_j \right) = - \sum_{\langle j 
\rangle} I_{i,j} =I_i \ ,
\end{eqnarray}
where $I_i = I \left( \delta_{i,x} - \delta_{i,x^\prime} \right)$ and the summations 
extend over the nearest neighbors of $i$.

Alternatively to Eq.~(\ref{powerInTermsOfV}) the power can by rewritten in terms of the currents as
\begin{eqnarray}
\label{powerInTermsOfI}
P=  R (x ,x^\prime) I^2 = \sum_{\langle i,j \rangle} \rho_{i,j} I_{i,j}^2 =  P 
\left( \left\{ I \right\} \right) \ ,
\end{eqnarray}
with $\left\{ I \right\}$ denoting the set of currents flowing through the 
individual bonds and $\rho_{i,j} = \sigma_{i,j}^{-1}$. Obviously the cluster may contain closed loops as subnetworks. Suppose there are currents $\left\{ I^{(l)} \right\}$ circulating independently around a complete set of independent closed loops. Then the power is not only a function of $I$ but also of the set of loop currents. The potential drop around closed loops is zero. This gives rise to the variation principle
\begin{eqnarray}
\label{variationPrinciple2}
\frac{\partial}{\partial I^{(l)}} P \left( \left\{ I^{(l)} \right\} , I \right) 
= 0 \ .
\end{eqnarray} 
Eq.~(\ref{variationPrinciple2}) may be used to eliminate the loop currents and 
thus provides us with a method to determine the total resistance of the backbone 
via Eq.~(\ref{powerInTermsOfI}).

Since the resistance of the backbone depends on the configurations $C$ of the randomly occupied bonds, one introduces an average $\langle \cdots \rangle_C$ over these configurations. It is important to recognize that the resistance between disconnected sites is infinite. Therefore one considers only those sites $x$ and $x^\prime$ known to be on the same cluster. Practically this is done by introducing the indicator function $\chi (x ,x^\prime)$ which, for a given configuration $C$, is unity if $x$ and $x^\prime$ are connected and zero otherwise. Then the $n$th moment of the resistance $R$ with respect to the average $\langle \cdots \rangle_C$ subject to $x$ and $x^\prime$ being on the same 
cluster is given by
\begin{eqnarray}
M_R^{(n)} = \left\langle \chi (x ,x^\prime) R (x ,x^\prime )^n \right\rangle_C / \left\langle \chi (x ,x^\prime) \right\rangle_C \ . 
\end{eqnarray}

\subsection{Generating function}
Our aim is to determine the average resistance $M_R = M_R^{(1)}$. Hence our task is twofold: we need to solve the set of Kirchhoff's equations~(\ref{cirquitEquations}) and to perform the average over all configurations of the diluted lattice. It can be accomplished by employing the replica technique\cite{stephen_78}. The voltages are replicated $D$-fold: $V_x \to \vec{V_x} = \left( V_x^{(1)}, \ldots , V_x^{(D)} \right)$. One introduces
\begin{eqnarray}
\psi_{\vec{\lambda}}(x) = \exp \left( i \vec{\lambda} \cdot \vec{V}_x \right) \ ,
\end{eqnarray}
where $\vec{\lambda} \cdot \vec{V}_x = \sum_{\alpha} \lambda^{(\alpha )} V_x^{(\alpha )}$ and $\vec{\lambda} \neq \vec{0}$. The corresponding correlation functions
\begin{eqnarray}
G \left( x, x^\prime ;\vec{\lambda} \right) = \left\langle \psi_{\vec{\lambda}}(x)\psi_{-\vec{\lambda}}(x^\prime) \right\rangle_{\mbox{\scriptsize{rep}}}
\end{eqnarray} 
are defined as
\begin{eqnarray}
\label{erzeugendeFunktion}
G \left( x, x^\prime ;\vec{\lambda} \right) = \lim_{D \to 0} \left\langle Z^{-D} \int \prod_j \prod_{\alpha =1}^D dV_j^\alpha \exp \left( -\frac{1}{2} P \left( \left\{ \vec{V} \right\} \right) + \frac{i\omega}{2} \sum_i \vec{V}^2_i + i \vec{\lambda} \cdot \left( \vec{V}_x  - \vec{V}_{x^\prime} \right) \right) \right\rangle_C \ .
\end{eqnarray}
Here $P \left( \left\{ \vec{V} \right\} \right) = \sum_{i,j,\alpha} \sigma_{i,j} \left( V_i^{(\alpha)} - V_j^{(\alpha)}\right)^{2}$ and $Z$ is the normalization
\begin{eqnarray}
\label{norm}
Z = \int \prod_{i} dV_{i} \exp \left[ -\frac{1}{2} P \left( \left\{ V \right\} \right) + \frac{i\omega}{2} \sum_i V^2_i \right] \ .
\end{eqnarray}
Note that we have introduced an additional power term $\frac{i\omega}{2} \sum_i V^2_i$. This is necessary to give the integrals in Eqs.~(\ref{erzeugendeFunktion}) and (\ref{norm}) a well defined meaning. Without this term the integrands depend only on voltage differences and the integrals are divergent. Physically the new term corresponds to grounding each lattice site by a capacitor of unit capacity. The original situation may be restored by taking the limit of vanishing frequency, $\omega \to 0$.

The integrations in Eq.~(\ref{erzeugendeFunktion}) can be carried out by employing the saddle point method. Since the integrations are Gaussian the saddle point method is exact in this case. The saddle point equation is identical to the variation principle stated in Eq.~(\ref{variationPrinciple1}). Thus the maximum of the integrand is determined by the solution of Kirchhoff's equations (\ref{cirquitEquations}) and 
\begin{eqnarray}
\label{genFkt}
G \left( x, x^\prime ;\vec{\lambda} \right) = \left\langle \exp \left( - \frac{\vec{\lambda}^2}{2} R \left( x,x^\prime \right) \right) \right\rangle_C \ ,
\end{eqnarray}
up to an unimportant multiplicative constant which goes to one in the limit $D \to 0$. Taylor expansion of the right hand side of Eq.~(\ref{genFkt}) about  
$\lambda^{(\alpha )} = 0$ leads to
\begin{eqnarray}
\label{expOfMomGenFkt}
G \left( x, x^\prime ;\vec{\lambda} \right) = \left\langle \chi( x, x^\prime) \right\rangle_C 
\left\{ 1 - \frac{\vec{\lambda}^2}{2} M_R^{(1)} \left( x, 
x^\prime \right) + \cdots + \frac{1}{k!} \left( - \frac{\vec{\lambda}^2}{2} \right)^k M_R^{(k)} \left( x, x^\prime \right) + \cdots \right\} \ .
\end{eqnarray}
Hence the correlation function $G$ can be exploited as a generating function for the 
moments $M_R^{(n)}$ given by
\begin{eqnarray}
\label{ableitungsvorschrift}
\left\langle \chi( x, x^\prime) \right\rangle_C M_R^{(n)} \left( x, x^\prime \right) = \frac{\partial^n}{\partial \left( - \vec{\lambda}^2 /2 \right)^n} G \left( x, x^\prime ;\vec{\lambda} \right) 
\Bigg|_{\vec{\lambda} = \vec{0}} \ .
\end{eqnarray}

\subsection{Field theoretic Hamiltonian}
Since infinite voltage drops between different clusters may occur, it is not guaranteed that $Z$ stays finite, i.e., the limit $\lim_{D\to 0}Z^D$ is not well defined. Moreover, $\vec{\lambda} = \vec{0}$ has to be excluded properly. Both problems can be handled by resorting to a lattice regularization of the integrals in Eqs.~(\ref{erzeugendeFunktion}) and (\ref{norm}). One switches to voltage variables $\vec{\theta}= \Delta \theta \vec{k}$ taking discrete values on a $D$-dimensional torus, i.e.\ $\vec{k}$ is chosen to be an 
$D$-dimensional integer with $-M < k^{(\alpha)} \leq M$ and $k^{(\alpha 
)}=k^{(\alpha )} \mbox{mod} (2M)$. $\Delta \theta = \theta_M / M$ is the gap between successive voltages and $\theta_M$ is the voltage cutoff. In this discrete picture there are $(2M)^D-1$ independent state variables per lattice site and one can introduce the Potts spins\cite{Zia_Wallace_75} 
\begin{eqnarray}
\Phi_{\vec{\theta}} \left( x \right) = (2M)^{-D} \sum_{\vec{\lambda} \neq 
\vec{0}} \exp \left( i \vec{\lambda} \cdot \vec{\theta} \right) 
\psi_{\vec{\lambda}} (x) = \delta_{\vec{\theta}, \vec{\theta}_{x}} - (2M)^{-D} 
\end{eqnarray}
subject to the condition $\sum_{\vec{\theta }} \Phi_{\vec{\theta}} \left( x 
\right) = 0$.

Now we revisit Eq.~(\ref{erzeugendeFunktion}). Carrying out the average over the diluted lattice configurations provides us with the weight $\exp (-H_{\mbox{\scriptsize{rep}}})$ of the average $\langle ... \rangle_{\mbox{\scriptsize{rep}}}$:
\begin{eqnarray}
H_{\mbox{\scriptsize{rep}}} &=& - \ln \left\langle  \exp \left( - \frac{1}{2} P + \frac{i\omega}{2} \sum_i \vec{\theta}_i^2 \right) \right\rangle_C 
\nonumber \\
&=& - \sum_{i ,j} \ln \left\langle  \exp \left( - \frac{1}{2} \sigma_{i ,j} \left( \vec{\theta}_{i} - \vec{\theta}_{j} \right)^2 \right)  \right\rangle_C + \frac{i\omega}{2} \sum_i \vec{\theta}_i^2 \ .
\end{eqnarray}
By dropping a constant term $N_{\mathcal B \mathnormal} \ln (1-p) + N_{\mathcal S \mathnormal} \ln (1-p_{\mathcal S \mathnormal})$ with $N_{\mathcal S \mathnormal}$ being the number of bonds between sites on the surface of the undiluted lattice and $N_{\mathcal B \mathnormal}$ being the corresponding quantity related to remaining bonds we obtain
\begin{eqnarray}
\label{HmitSundB}
H_{\mbox{\scriptsize{rep}}} &=& - \sum_{i \in \mathcal B \mathnormal ,j \in \mathcal B \mathnormal \cup \mathcal S \mathnormal } K_{\mathcal B \mathnormal} \left( \vec{\theta}_{i} - \vec{\theta}_{j} \right) - \sum_{i \in \mathcal S \mathnormal ,j \in \mathcal S \mathnormal } K_{\mathcal S \mathnormal} \left( \vec{\theta}_{i} - \vec{\theta}_{j} \right) + \sum_i h \left( \vec{\theta}_{i} \right) 
\nonumber \\
&=& - \sum_{i \in \mathcal B \mathnormal ,j \in \mathcal B \mathnormal \cup \mathcal S \mathnormal } \sum_{\vec{\theta},\vec{\theta}^\prime} K_{\mathcal B \mathnormal} \left( \vec{\theta} - \vec{\theta}^\prime \right) \Phi_{\vec{\theta}} \left( i \right) \Phi_{\vec{\theta}^\prime} \left( j \right) 
- \sum_{i \in \mathcal S \mathnormal ,j \in \mathcal S \mathnormal} \sum_{\vec{\theta},\vec{\theta}^\prime} K_{\mathcal S \mathnormal} \left( \vec{\theta} - \vec{\theta}^\prime \right) \Phi_{\vec{\theta}} \left( i \right) \Phi_{\vec{\theta}^\prime} \left( j \right)
\nonumber \\
& &+ \sum_i \sum_{\vec{\theta}} h \left( \vec{\theta} \right) \Phi_{\vec{\theta}} \left( i \right) \ ,
\end{eqnarray}
where
\begin{eqnarray}
K_{\mathcal B \mathnormal} \left( \vec{\theta} \right) = \ln \left\{ 1 + \frac{p}{1-p} \exp \left( - \frac{1}{2} \sigma \vec{\theta} \cdot \vec{\theta} \right) \right\} 
\end{eqnarray}
and
\begin{eqnarray}
K_{\mathcal S \mathnormal} \left( \vec{\theta} \right) = \ln \left\{ 1 + \frac{p_{\mathcal S \mathnormal}}{1-p_{\mathcal S \mathnormal}} \exp \left( - \frac{1}{2} \sigma_{\mathcal S \mathnormal} \vec{\theta} \cdot \vec{\theta} \right) \right\} \ . 
\end{eqnarray}
The summations in Eq.~(\ref{HmitSundB}) run over nearest neighbor pairs subject to the specified conditions. Note that $K_{\mathcal B \mathnormal} \left( \vec{\theta} \right)$ and $K_{\mathcal S \mathnormal} \left( \vec{\theta} \right)$ are exponentially decreasing functions in replica space with a decay rate proportional to $\sigma^{-1}$ and $\sigma_{\mathcal S \mathnormal}^{-1}$ respectively. The Hamiltonian $H_{\mbox{\scriptsize{rep}}}$ describes interaction of Potts spins in a semi-infinite system with an external one site potential $ h \left( \vec{\theta} \right)$. The interaction is rotationally and translationally invariant in replica space. For large $\sigma$ and $\sigma_{\mathcal S \mathnormal}$ the interaction is short ranged not only in real but also in replica space.  Moreover, the interaction potential $K_{\mathcal B \mathnormal} \left( \vec{\theta} \right)$ is an analytic function of $\vec{\theta}^2$. Thus the Fourier transform 
\begin{eqnarray}
\widetilde{K}_{\mathcal B \mathnormal} \left( \vec{\lambda} \right) = - \frac{1}{(2M)^D} \sum_{\vec{\theta}} \exp \left( -i \vec{\lambda} \cdot \vec{\theta} \right) \ln \left( 1 + \frac{p}{1-p} \exp \left( - \frac{1}{2} \sigma \vec{\theta} \cdot \vec{\theta} \right) \right)
\end{eqnarray}
can be Taylor expanded as 
\begin{eqnarray}
\label{taylorExpB}
\widetilde{K}_{\mathcal B \mathnormal} \left( \vec{\lambda} \right) = \tau + \sum_{p=1}^{\infty} w_p \left( \vec{\lambda}^2 \right)^p \ ,
\end{eqnarray}
with $\tau$ and $w_p \sim \sigma^{-p}$ being expansion coefficients. Analogously one obtains for the Fourier transform of $K_{\mathcal S \mathnormal}$:
\begin{eqnarray}
\label{taylorExpS}
\widetilde{K}_{\mathcal S \mathnormal} \left( \vec{\lambda} \right) = c + \sum_{p=1}^{\infty} w_{\mathcal S \mathnormal ,p} \left( \vec{\lambda}^2 \right)^p \ ,
\end{eqnarray}
with coefficients $c$ and $w_{\mathcal S \mathnormal ,p} \sim \sigma_{\mathcal S \mathnormal}^{-p}$.
 
In the limit of perfect transport, $\sigma \to \infty$ and $\sigma_{\mathcal S \mathnormal} \to \infty$, $K_{\mathcal B \mathnormal} \left( \vec{\theta} \right)$ and $K_{\mathcal S \mathnormal} \left( \vec{\theta} \right)$ go to their local limits $K_{\mathcal B \mathnormal} \left( \vec{\theta} \right) = K_{\mathcal B \mathnormal} \delta_{\vec{\theta}, \vec{0}}$ and $K_{\mathcal S \mathnormal} \left( \vec{\theta} \right) = K_{\mathcal S \mathnormal} \delta_{\vec{\theta}, \vec{0}}$, with $K_{\mathcal B \mathnormal}$ and $K_{\mathcal S \mathnormal}$ being positive constants. The interaction part of the Hamiltonian reduces to
\begin{eqnarray}
\label{interActionHamil}
H^{\mbox{\scriptsize{int}}}_{\mbox{\scriptsize{rep}}} = - K_{\mathcal B \mathnormal} \sum_{i \in \mathcal B \mathnormal ,j \in \mathcal B \mathnormal \cup \mathcal S \mathnormal }  \sum_{\vec{\theta}} \Phi_{\vec{\theta}} \left( i \right) \Phi_{\vec{\theta}} \left( j \right) - K_{\mathcal S \mathnormal} \sum_{i \in \mathcal B \mathnormal ,j \in \mathcal B \mathnormal}  \sum_{\vec{\theta}} \Phi_{\vec{\theta}} \left( i \right) \Phi_{\vec{\theta}} \left( j \right) \ .
\end{eqnarray}
This represents nothing more than the semi-infinite $\left( 2M \right)^D$ states Potts model which is invariant against all $\left( 2M \right)^D !$ permutations of the Potts spins $\Phi_{\vec{\theta}}$. If $\sigma^{-1} \neq 0$ and $\sigma_{\mathcal S \mathnormal}^{-1} \neq 0$, this $S_{\left( 2M \right)^D}$ symmetry is lost in favor of the short range interactions.

We proceed with the usual coarse graining step and replace the Potts spins 
$\Phi_{\vec{\theta}} \left( x \right)$ by order parameter fields $\varphi 
\left( {\rm{\bf x}} ,\vec{\theta} \right)$ which inherit the constraint $\sum_{\vec{\theta}} \varphi \left( {\rm{\bf x}} ,\vec{\theta} \right) = 0$. We model the corresponding field theoretic Hamiltonian $\mathcal H \mathnormal$ in the spirit of Landau as a mesoscopic free energy from local monomials of the order parameter field and its gradients in real and replica space. The gradient expansion is justified since the interaction is short ranged in both spaces. Purely local terms in replica space have to respect the full $S_{\left( 2M \right)^D}$ Potts symmetry. After these remarks we write down the Landau-Ginzburg-Wilson type Hamiltonian
\begin{mathletters}
\label{hamiltonfunktion}
\begin{eqnarray}
\mathcal H \mathnormal = {\mathcal H \mathnormal}_{\mathcal B \mathnormal} + {\mathcal H \mathnormal}_{\mathcal S \mathnormal}\ ,
\end{eqnarray}
with the bulk contribution   
\begin{eqnarray}
\label{bulkHamiltonian}
{\mathcal H \mathnormal}_{\mathcal B \mathnormal} = \int_{V} d^dx \sum_{\vec{\theta}} \left\{ \frac{1}{2} \varphi \left( {\rm{\bf x}} , \vec{\theta} \right) K_{\mathcal B \mathnormal} \left( \Delta ,\Delta_{\vec{\theta}} \right) \varphi \left( {\rm{\bf x}} , \vec{\theta} \right) + \frac{g}{6}\varphi \left( {\rm{\bf x}} , \vec{\theta} \right)^3 \right\} \ , 
\end{eqnarray}
and the surface part
\begin{eqnarray}
\label{surfaceHamiltonian}
{\mathcal H \mathnormal}_{\mathcal S \mathnormal} = \int_{\partial V} d^{d-1}x_{\|} \sum_{\vec{\theta}} \left\{ \frac{1}{2} \varphi_{\mathcal S \mathnormal} \left( {\rm{\bf x}}_{\|} , \vec{\theta} \right) K_{\mathcal S \mathnormal} \left( \Delta_{\vec{\theta}} \right) \varphi_{\mathcal S \mathnormal} \left( {\rm{\bf x}}_{\|} , \vec{\theta} \right) \right\} \ , 
\end{eqnarray}
where
\begin{eqnarray}
K_{\mathcal{B}} \left( \Delta ,\Delta_{\vec{\theta}} \right) = \tau - \Delta - w \Delta_{\vec{\theta}}
\end{eqnarray}
and
\begin{eqnarray}
K_{\mathcal{S}} \left( \Delta_{\vec{\theta}} \right) = c - w_{{\mathcal{S}}} \Delta_{\vec{\theta}} \ .
\end{eqnarray}
\end{mathletters}
The integration in Eq.~(\ref{bulkHamiltonian}) extends over the half-space $V = \big\{ {\bf x} = ( {\bf x}_\parallel , z ) \big| \, {\bf x}_\parallel \in {\rm{\bf R}}^{d-1}, z \geq 0 \big\}$ whereas the integration in Eq.~(\ref{surfaceHamiltonian}) extends over the corresponding boundary $\partial V$. Terms of higher order in the fields have been neglected in Eq.~(\ref{bulkHamiltonian}) and Eq.~(\ref{surfaceHamiltonian}) since they turn out to be irrelevant in the renormalization group sense. Moreover, a term of the form $\varphi_{\mathcal S \mathnormal} \partial_n \varphi = \lim_{z \to 0^{+}} \varphi_{\mathcal S \mathnormal} \partial_z \varphi$ has been neglected in Eq.~(\ref{surfaceHamiltonian}) though it is marginal on dimensional grounds. Such a term turns out to be redundant, cf.\ Ref.~\cite{diehl_86}. $\tau$, $c$, $w$, and $w_{\mathcal S \mathnormal}$ are now coarse grained analogues of the original coefficients. Note that, upon setting $w=w_{\mathcal S \mathnormal}=0$, $\mathcal H \mathnormal$ reduces to the usual semi-infinite Potts model Hamiltonian as studied by Diehl and Lam\cite{diehl_lam_89}.

\subsection{Irrelevance of $w_{\mathcal{S}}$}
\label{relevannceOfWs}
Here we discuss the relevance of $w_{\mathcal S \mathnormal}$ in the renormalization group sense. Let $P$ denote the set of parameters $\left\{ \tau , w, c, w_{\mathcal S \mathnormal} \right\}$ and $b$ some scaling factor for the voltage variable: $\vec{\theta} \to b \vec{\theta}$. By substitution of $\varphi \left( {\rm{\bf x}} , \vec{\theta} \right) = \varphi^\prime \left( {\rm{\bf x}} , b\vec{\theta} \right)$ and $\varphi_{\mathcal S \mathnormal} \left( {\rm{\bf x}} , \vec{\theta} \right) = \varphi_{\mathcal S \mathnormal}^\prime \left( {\rm{\bf x}}_{\|} , b\vec{\theta} \right)$ into the Hamiltonian we get
\begin{eqnarray}
\label{semiScaling1}
\lefteqn{ {\mathcal H \mathnormal} \left[ \varphi^\prime \left( {\rm{\bf x}} , b \vec{\theta} \right) , \varphi_{\mathcal S \mathnormal}^\prime \left( {\rm{\bf x}}_{\|} , b \vec{\theta} \right) , P \right] }
\nonumber \\
&=& \int_V d^dx \sum_{\vec{\theta}} \left\{ \frac{1}{2} \varphi^\prime \left( {\rm{\bf x}} , b \vec{\theta} \right) K_{\mathcal B \mathnormal} \left( \Delta ,\Delta_{\vec{\theta}} \right) \varphi^\prime \left( {\rm{\bf x}} , b \vec{\theta} \right) + \frac{g}{6}\varphi^{\prime } \left( {\rm{\bf x}} , b \vec{\theta} \right)^3 \right. \Bigg\} 
\nonumber \\
&+& \int_{\partial V} d^{d-1}x_{\|} \sum_{\vec{\theta}} \left\{ \frac{1}{2} \varphi_{\mathcal S \mathnormal}^\prime \left( {\rm{\bf x}}_{\|} , b \vec{\theta} \right) K_{\mathcal S \mathnormal} \left( \Delta_{\vec{\theta}} \right) \varphi_{\mathcal S \mathnormal}^\prime \left( {\rm{\bf x}}_{\|} , b \vec{\theta} \right) \right\}
\end{eqnarray}
Renaming the scaled voltage variables, $\vec{\theta}^\prime = b \vec{\theta}$, leads to
\begin{eqnarray}
\label{semiScaling2}
\lefteqn{ {\mathcal H \mathnormal} \left[ \varphi^\prime \left( {\rm{\bf x}} , \vec{\theta}^\prime \right) , \varphi_{\mathcal S \mathnormal}^\prime \left( {\rm{\bf x}}_{\|} , \vec{\theta}^\prime \right) , P \right] }
\nonumber \\
&=& \int_V d^dx \sum_{\vec{\theta}^\prime} \left\{ \frac{1}{2} \varphi^\prime \left( {\rm{\bf x}} , \vec{\theta}^\prime \right) K_{\mathcal B \mathnormal} \left( \Delta , b^2 \Delta_{\vec{\theta}^\prime} \right) \varphi^\prime \left( {\rm{\bf x}} , \vec{\theta}^\prime \right) + \frac{g}{6}\varphi^{\prime } \left( {\rm{\bf x}} , \vec{\theta}^\prime \right)^3 \right. \Bigg\} 
\nonumber \\
&+& \int_{\partial V} d^{d-1}x_{\|} \sum_{\vec{\theta}^\prime} \left\{ \frac{1}{2} \varphi_{\mathcal S \mathnormal}^\prime \left( {\rm{\bf x}}_{\|} , \vec{\theta}^\prime \right) K_{\mathcal S \mathnormal} \left( b^2 \Delta_{\vec{\theta}^\prime} \right) \varphi_{\mathcal S \mathnormal}^\prime \left( {\rm{\bf x}}_{\|} , \vec{\theta}^\prime \right) \right\}
\end{eqnarray}
Clearly a scaling of the voltage variable results in a scaling of the voltage cutoff: $\theta_M \to b \theta_M$. However, by taking the limit $D \to 0$ before $\theta_M \to \infty$ the dependence of the theory on the cutoff drops out. We can identify $\vec{\theta}^\prime$ and $\vec{\theta}$ and thus 
\begin{eqnarray}
\label{semiRelForH}
{\mathcal H \mathnormal} \left[ \varphi \left( {\rm{\bf x}} , b \vec{\theta} \right) , \varphi_{\mathcal S \mathnormal} \left( {\rm{\bf x}}_{\|} , b \vec{\theta} \right) , P \right] = {\mathcal H \mathnormal} \left[ \varphi \left( {\rm{\bf x}} , \vec{\theta} \right) , \varphi_{\mathcal S \mathnormal} \left( {\rm{\bf x}}_{\|} , \vec{\theta} \right) , P^\prime \right] \ ,
\end{eqnarray}
where $P^\prime  = \left\{ \tau , b^2 w, c, b^2 w_{\mathcal S \mathnormal} \right\}$. Equation~(\ref{semiRelForH}) implies for the connected correlation functions
\begin{eqnarray}
G_{N,M} \left( \left\{ {\rm{\bf x}}, {\rm{\bf x}}_{\|}, \vec{\theta} \right\} ; \tau ,w ,c, w_{\mathcal S \mathnormal} \right) = \lim_{D \to 0} \left\langle \prod_{i=1}^{N} \varphi \left( {\rm{\bf x}}_{i} , \vec{\theta}_{i} \right) \prod_{j=1}^{M} \varphi_{\mathcal S \mathnormal} \left( {\rm{\bf x}}_{\| j} , \vec{\theta}_{j} \right) \right\rangle^{\mbox{\scriptsize{conn}}}_{\mathcal H \mathnormal}
\end{eqnarray}
that
\begin{eqnarray}
G_{N,M} \left( \left\{ {\rm{\bf x}}, {\rm{\bf x}}_{\|}, \vec{\theta} \right\} ; \tau ,w ,c, w_{\mathcal S \mathnormal} \right) = G_{N,M} \left( \left\{ {\rm{\bf x}}, {\rm{\bf x}}_{\|}, b \vec{\theta} \right\} ; \tau ,b^2 w ,c, b^2 w_{\mathcal S \mathnormal} \right) \ .
\end{eqnarray}
Note that ${\rm{\bf x}}$ in $G_{N,M}$ refers to a point located off the surface, i.e, ${\rm{\bf x}} = \left( {\rm{\bf x}}_\| , z \right)$ with $z>0$. In other quantities, however, ${\rm{\bf x}}$ may refer to an arbitrary point. The two-point correlation functions $G_{N,M}$ with $N+M=2$ are nothing more than replica space Fourier transforms of  $ \left\langle \psi_{\vec{\lambda}}({\rm{\bf x}}) \psi_{-\vec{\lambda}}({\rm{\bf x}}) \right\rangle_{\mathcal H \mathnormal}$. We deduce from Eq.~(\ref{expOfMomGenFkt}) that
\begin{eqnarray}
\vec{\lambda}^2 M_R \left( \left( {\rm{\bf x}} , {\rm{\bf x}}^\prime \right) ;\tau , w, c, w_{\mathcal S \mathnormal} \right) = \left( b^{-1} \vec{\lambda}  \right)^2 M_R \left( \left( {\rm{\bf x}} , {\rm{\bf x}}^\prime \right) ;\tau , b^2 w , c, b^2 w_{\mathcal S \mathnormal} \right) \ .
\end{eqnarray}
The freedom to choose $b$ has not been exploited yet. With the choice $b^2 = w^{-1}$ the previous scaling relation turns into
\begin{eqnarray}
\label{scaling_w}
M_R \left( \left( {\rm{\bf x}} , {\rm{\bf x}}^\prime \right) ;\tau , , w, c, w_{\mathcal S \mathnormal} \right) = w f \left( \left( {\rm{\bf x}} , {\rm{\bf x}}^\prime \right) ;\tau , c , \frac{w_{\mathcal S \mathnormal}}{w} \right) \ ,
\end{eqnarray}
where $f$ is a scaling function. Dimensional analysis of the Hamiltonian shows that $w\vec{\lambda}^2 \sim \mu^2$ and $w_{\mathcal S \mathnormal} \vec{\lambda}^2 \sim \mu^1$, where $\mu$ is the usual inverse length scale. Thus $w_{\mathcal S \mathnormal} /w \sim \mu^{-1}$, i.e., $w_{\mathcal S \mathnormal}$ is an irrelevant coupling.

\section{Renormalization group analysis}
\label{rga}
The three types of transition occurring at bulk criticality $\tau = 0$, namely the ordinary, the special, and the extraordinary transition, are described by renormalization group (RG) fixed points where $c$ takes the value $c^\ast_{\mbox{\scriptsize{ord}}} = \infty$, $c^\ast_{\mbox{\scriptsize{sp}}} = 0$, and $c^\ast_{\mbox{\scriptsize{ex}}} = - \infty$  respectively (cf.\ Sec.~III\,C\,3 of Rev.~\cite{diehl_86} and references therein). In the remainder of this paper we focus on the special and the ordinary transition. This section provides an outline of our renormalization group improved perturbation calculation.

\subsection{Gaussian propagator}
Now we determine the Gaussian propagator for our model. For this purpose the irrelevant term has to be discarded from the Hamiltonian. Then the saddle point solution of
\begin{eqnarray}
\int \mathcal D \mathnormal \varphi \exp \left( -\mathcal H \mathnormal \right) \ ,
\end{eqnarray}
where $\mathcal D \mathnormal \varphi$ indicates an integration over the set of variables $\left\{ \varphi \left( {\rm{\bf x}} , 
\vec{\theta} \right) \right\}$ for all ${\rm{\bf x}}$ and $\vec{\theta}$, is determined by the so-called equations of motion
\begin{mathletters}
\label{eqMo}
\begin{eqnarray}
\label{eqMo1}
\tau \varphi - \Delta \varphi - w \Delta_{\vec{\theta}} \varphi + \frac{g}{3} \varphi^2 = 0
\end{eqnarray}
and
\begin{eqnarray}
\label{eqMo2}
\varphi_{\mathcal S \mathnormal} + \partial_n \varphi_{\mathcal S \mathnormal} = 0 \ .
\end{eqnarray}
\end{mathletters}
In Eq.~(\ref{eqMo2}) $\partial_n \varphi_{\mathcal S \mathnormal} \left( {\rm{\bf x}}_{\|} , \vec{\theta} \right)$ denotes the normal derivative $\lim_{z\to 0^{+}} \partial_z \varphi \left( {\rm{\bf x}}, \vec{\theta} \right)$.
As a consequence of Eq.~(\ref{eqMo}), the Fourier transform $G_{{\rm{\bf p}},\vec{\lambda}} \left( z, z^\prime \right)$ of the Gaussian propagator $G \left( {\rm{\bf x}}_{\|} - {\rm{\bf x}}_{\|}^\prime ,z,z^\prime , \vec{\theta} - \vec{\theta}^\prime \right)$ has to satisfy
\begin{mathletters}
\label{bestimmungsGl}
\begin{eqnarray}
\left[ \tau + {\rm{\bf p}}^2 + w \vec{\lambda}^2 - \partial_z^2 \right] G_{{\rm{\bf p}},\vec{\lambda}} \left( z, z^\prime \right) = \delta \left( z - z^\prime \right)
\end{eqnarray}
with the boundary conditions
\begin{eqnarray}
c \, G_{{\rm{\bf p}},\vec{\lambda}} \left( z, 0 \right) = \partial_{z^\prime} G_{{\rm{\bf p}},\vec{\lambda}} \left( z, z^\prime \right) \Big|_{z^\prime =0}
\end{eqnarray}
and
\begin{eqnarray}
G_{{\rm{\bf p}},\vec{0}} \left( z, z^\prime \right) = 0 \ .
\end{eqnarray}
\end{mathletters}
Equations~(\ref{bestimmungsGl}) can by solved as described in Ref.~\cite{diehl_86}, and one obtains
\begin{eqnarray}
\label{semiInfProp}
G_{{\rm{\bf p}},\vec{\lambda}} \left( z, z^\prime \right) = \frac{1}{2 \kappa_{\vec{\lambda}}} \left[ \exp \left( - \kappa_{\vec{\lambda}} \left| z - z^\prime \right| \right) + \frac{\kappa_{\vec{\lambda}} - c}{\kappa_{\vec{\lambda}} + c} \exp \left( - \kappa_{\vec{\lambda}} \left( z + z^\prime \right) \right) \right] \left( 1 - \delta_{\vec{\lambda},\vec{0}} \right) \ ,
\end{eqnarray}
where
\begin{eqnarray}
\kappa_{\vec{\lambda}} = \left( \tau + {\rm{\bf p}}^2 + w \vec{\lambda}^2 \right)^{1/2} \ .
\end{eqnarray}
As in our previous work on RRN~\cite{stenull_janssen_oerding_99,janssen_stenull_oerding_99,janssen_stenull_99,stenull_janssen_2000a,stenull_janssen_2000b}, this principal propagator may be viewed as being composed of a conducting and an insulating part:
\begin{eqnarray}
G_{{\rm{\bf p}},\vec{\lambda}} \left( z, z^\prime \right) =  G^{\mbox{\scriptsize{cond}}}_{{\rm{\bf p}},\vec{\lambda}} \left( z, z^\prime \right) - G^{\mbox{\scriptsize{ins}}}_{{\rm{\bf p}}} \left( z, z^\prime \right) \ ,
\end{eqnarray}
where
\begin{eqnarray}
 G^{\mbox{\scriptsize{cond}}}_{{\rm{\bf p}},\vec{\lambda}} \left( z, z^\prime \right) = \frac{1}{2 \kappa_{\vec{\lambda}}} \left[ \exp \left( - \kappa_{\vec{\lambda}} \left| z - z^\prime \right| \right) + \frac{\kappa_{\vec{\lambda}} - c}{\kappa_{\vec{\lambda}} + c} \exp \left( - \kappa_{\vec{\lambda}} \left( z + z^\prime \right) \right) \right]
\end{eqnarray}
and 
\begin{eqnarray}
 G^{\mbox{\scriptsize{ins}}}_{{\rm{\bf p}}} \left( z, z^\prime \right) = \frac{1}{2 \kappa_{\vec{0}}} \left[ \exp \left( - \kappa_{\vec{0}} \left| z - z^\prime \right| \right) + \frac{\kappa_{\vec{0}} - c }{\kappa_{\vec{0}} + c } \exp \left( - \kappa_{\vec{0}} \left( z + z^\prime \right) \right) \right] \delta_{\vec{\lambda},\vec{0}} \ .
\end{eqnarray}

\subsection{Special transition}
\label{specialTrans}
Here we sketch   our renormalization group improved perturbation calculation for the special transition. The ordinary transition is relegated to the next subsection.

\begin{figure}[h]
\epsfxsize=2.3cm
\begin{eqnarray*}
\raisebox{-10mm}{\epsffile{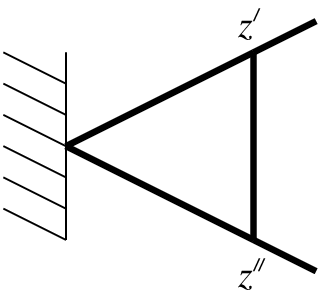}} = \ \raisebox{-11mm}{\epsffile{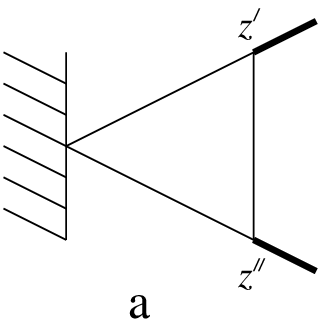}} - \ \raisebox{-11mm}{\epsffile{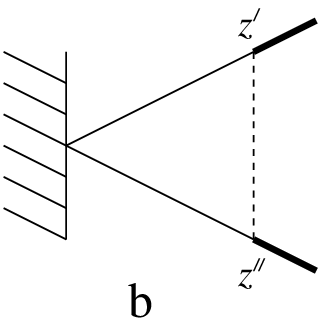}}
\end{eqnarray*}
\caption[]{\label{specialDiagrams}Ultraviolet divergent Feynman diagrams contributing to the renormalization of $\mathcal{O}_{\mathcal S \mathnormal}$ at the special transition. The bold lines represent principal propagators $G_{{\rm{\bf p}},\vec{\lambda}} \left( z, z^\prime \right)$. The light lines stand for conducting propagators $G^{\mbox{\scriptsize{cond}}}_{{\rm{\bf p}},\vec{\lambda}} \left( z, z^\prime \right)$ and the dashed lines represent insulating propagators $G^{\mbox{\scriptsize{ins}}}_{{\rm{\bf p}}} \left( z, z^\prime \right)$. Insertions of the surface operator ${\mathcal O \mathnormal}_{\mathcal S \mathnormal}$ are symbolized by the hatched ''surface''.}
\end{figure}

The constituting elements of our diagrammatic expansion are the vertex $-g$ and the propagator 
$G_{{\rm{\bf p}},\vec{\lambda}} \left( z, z^\prime \right)$. The decomposition of the principle propagator into $G^{\mbox{\scriptsize{cond}}}_{{\rm{\bf p}},\vec{\lambda}} \left( z, z^\prime \right)$ and $G^{\mbox{\scriptsize{ins}}}_{{\rm{\bf p}}} \left( z, z^\prime \right)$ allows for a schematic decomposition of the principle Feynman diagrams into sums of conducting diagrams consisting of conducting and insulating propagators (see Fig.~\ref{specialDiagrams} and Fig.~\ref{ordinaryDiagrams}). These conducting diagrams may be interpreted as being resistor networks themselves with conducting propagators corresponding to conductors and insulating propagators corresponding to open bonds. 

At the special transition $G^{\mbox{\scriptsize{cond}}}_{{\rm{\bf p}},\vec{\lambda}} \left( z, z^\prime \right)$ simplifies to the conducting Neumann propagator
\begin{eqnarray}
\label{NeumannProp}
G^{\mbox{\scriptsize{cond}},N}_{{\rm{\bf p}},\vec{\lambda}} \left( z, z^\prime \right) = \frac{1}{2 \kappa_{\vec{\lambda}} } \left[ \exp \left( - \kappa_{\vec{\lambda}} \left| z - z^\prime \right| \right) + \exp \left( - \kappa_{\vec{\lambda}} \left( z + z^\prime \right) \right) \right] \ .
\end{eqnarray}
It is convenient to parametrize the conducting Neumann propagator in terms of modified Bessel functions of the second kind\cite{gradshteyn_ryzhik}:
\begin{eqnarray}
G^{\mbox{\scriptsize{cond}},N}_{{\rm{\bf p}},\vec{\lambda}} \left( z, z^\prime \right) &=& 
\frac{1}{\left( 4 \pi \right)^{1/2}} \int_0^\infty \frac{ds}{s^{1/2}} \ \exp \left[ -s \left( \tau + {\rm{\bf p}}^2 + w \vec{\lambda}^2 \right) \right] 
\nonumber \\
&\times& \left\{ \exp \left[ - \frac{\left( z - z^\prime \right)^2}{4s} \right] +  \exp \left[ - \frac{\left( z + z^\prime \right)^2}{4s} \right] \right\} \ .
\end{eqnarray}
In this parametrization the parameters $s$ correspond to resistances and the replica variables $i\vec{\lambda}_i$ to currents. The replica currents are conserved in each vertex and we may write $\vec{\lambda}_i = \vec{\lambda}_i \left( \vec{\lambda} , \left\{ \vec{\lambda^{(l)}} \right\} \right)$, where $\vec{\lambda}$ is an external current and $\left\{ \vec{\lambda}^{(l)} \right\}$ denotes the set of independent loop currents.

As we have learned in Sec.~\ref{relevannceOfWs}, the surface coupling constant $w_{\mathcal S \mathnormal}$ is irrelevant. By setting it to zero, however, we would loose all information about its impact on the average resistance. To investigate this impact we work with insertions~\cite{amit_zinn-justin}. Here at the special transition we analyse correlation functions
\begin{eqnarray}
G_{N,M} \left( \left\{ {\rm{\bf x}}, {\rm{\bf x}}_{\|}, \vec{\theta} \right\} ; \tau ,w ,c \right)_{\mathcal{O}_{\mathcal{S}}} = \lim_{D \to 0} \left\langle {\mathcal{O}_{\mathcal{S}}} \prod_{i=1}^{N} \varphi \left( {\rm{\bf x}}_{i} , \vec{\theta}_{i} \right) \prod_{j=1}^{M} \varphi_{\mathcal S \mathnormal} \left( {\rm{\bf x}}_{\| j} , \vec{\theta}_{j} \right) \right\rangle^{\mbox{\scriptsize{conn}}}_{\mathcal H \mathnormal}
\end{eqnarray}
where the operator
\begin{eqnarray}
\label{operatorDef1}
{\mathcal{O}}_{{\mathcal{S}}} = \frac{w}{2} \int d^{d-1} x_\parallel \, \sum_{\vec{\theta}} \, \left( \nabla_{\vec{\theta}} \varphi_{\mathcal S \mathnormal} \left( {\rm{\bf x}}_{\|} , \vec{\theta} \right) \right)^2
\end{eqnarray}
is inserted. Note by comparing Eqs.~(\ref{hamiltonfunktion}) and (\ref{operatorDef1}) that ${\mathcal{O}}_{{\mathcal{S}}}$ is associated with a coupling constant $v_{{\mathcal{S}}} = w_{{\mathcal{S}}}/w$.

To see how the insertion procedure works in detail we now consider diagram a (see Fig.~\ref{specialDiagrams}). Diagram a stands for
\begin{eqnarray}
\label{surfaceDiagram_a}
\mbox{a} = - g^2 w \sum_{\vec{\lambda}^{(l)}} \int_{{\rm{\bf p}}} \vec{\lambda}^{(l)2} G^{\mbox{\scriptsize{cond}},N}_{{\rm{\bf p}},\vec{\lambda}^{(l)}} \left( z^\prime, 0 \right) G^{\mbox{\scriptsize{cond}},N}_{{\rm{\bf p}},\vec{\lambda}^{(l)}} \left( 0, z^{\prime \prime} \right) G^{\mbox{\scriptsize{cond}},N}_{{\rm{\bf p}},\vec{\lambda} + \vec{\lambda}^{(l)}} \left( z^\prime, z^{\prime \prime} \right) \ ,
\end{eqnarray}
where $\int_{{\rm{\bf p}}}$ is an abbreviation for $(2\pi )^{-(d-1)}\int d^{d-1}p$. The involved summation over the loop current reads
\begin{eqnarray}
\label{currentSum1}
- w \sum_{\vec{\lambda}^{(l)}} \vec{\lambda}^{(l)2} \exp \left[ w P \left( \vec{\lambda}, \vec{\lambda}^{(l)}\right) \right] \ .
\end{eqnarray}
We interpret $P \left( \vec{\lambda}, \vec{\lambda}^{(l)}\right) = - \left( s_2 + s_3 \right) \vec{\lambda}^{(l)2} - s_1 \left( \vec{\lambda} + \vec{\lambda}^{(l)} \right)^2$ as the power of the diagram. To evaluate the summation we employ the saddle point method. Note that the saddle point equation is nothing more than the variation principle stated in Eq.~(\ref{variationPrinciple2}). Thus, solving the saddle point equations is equivalent to determining the total resistance $R \left( \left\{ s_i \right\} \right)$ of a diagram, and the saddle point evaluation of (\ref{currentSum1}) yields
\begin{eqnarray}
\label{currentSum2}
\mbox{(\ref{currentSum1})} &=& - w \exp \left[ - R \left( s_1 , s_2 , s_3 \right) w \vec{\lambda}^2 \right] \int_{-\infty}^\infty d^d \lambda^{(l)} \ \left[ \vec{\lambda}^{(l)} - \frac{s_1}{s_1 + s_2 + s_3} \vec{\lambda} \right]^2
\nonumber \\
&\times& \exp \left[ - \left( s_1 + s_2 + s_3 \right) w \vec{\lambda}^{(l)2} \right] \ .
\end{eqnarray}
Here we have switched back to continuous currents. The remaining integration is Gaussian and therefore straightforward. In the limit $D\to 0$ we obtain upon Taylor expansion of $\exp \left[ - R \left( s_1 , s_2 , s_3 \right) w \vec{\lambda}^2 \right]$,
\begin{eqnarray}
\label{currentSum3}
\mbox{(\ref{currentSum2})} = - w \vec{\lambda}^2 \left( \frac{s_1}{s_1 + s_2 + s_3} \right)^2 + {\sl O} \left( \left( \vec{\lambda}^2 \right)^2 \right) \ .
\end{eqnarray}
Note that $i s_1 \vec{\lambda}/\left( s_1 + s_2 + s_3 \right)$ is the replica current flowing through the surface contact. Thus, we have an effective method to carry out the summations over the loop currents: essentialy we just need to determine the current flowing through the surface contact of the diagram. 

The ultraviolet divergent integrals occurring in the diagrammatic expansion can be regularized by the minimal subtraction method. We employ the following renormalization scheme:
\begin{mathletters}
\begin{eqnarray}
\varphi \to {\mathaccent"7017 \varphi} = Z^{1/2} \varphi \ ,&\quad&
\tau \to {\mathaccent"7017 \tau} = Z^{-1} Z_{\tau} \tau \ ,
\\
w \to {\mathaccent"7017 w} = Z^{-1} Z_w w \ , &\quad&
g \to {\mathaccent"7017 g} = Z^{-3/2} Z_u^{1/2} G_\epsilon^{-1/2} u^{1/2} \mu^{\epsilon /2}
\\
\varphi_{\mathcal S \mathnormal} \to {\mathaccent"7017 \varphi_{\mathcal S \mathnormal}} = Z_{1}^{1/2} Z^{1/2} \varphi_{\mathcal S \mathnormal} \ ,&\quad&
c \to {\mathaccent"7017 c} = Z_{1}^{-1}Z^{-1} Z_{c} c \ ,
\\
{\mathcal{O}}_{{\mathcal{S}}} \to \mathaccent"7017{{\mathcal{O}}}_{{\mathcal{S}}} &=& Z_{{\mathcal{O}}_{{\mathcal{S}}}} {\mathcal{O}}_{{\mathcal{S}}}\ ,
\end{eqnarray}
\end{mathletters}
where $G_\epsilon = (4\pi )^{-d/2}\Gamma (1 + \epsilon /2)$ with $\Gamma$ denoting the Gamma function. $Z$, $Z_{\tau}$, and $Z_{u}$ have been calculated to three-loop order by de Alcantara Bonfim {\it et al}\cite{alcantara_80}. Diehl and Lam\cite{diehl_lam_89} computed $Z_{1}$ and $Z_{c}$ to one loop order. Moreover, we gave a two-loop result for $Z_{w}$ in\cite{stenull_janssen_oerding_99}. Hence, it remains to determine $Z_{{\mathcal{O}}_{{\mathcal{S}}}}$. Our calculation which is sketched in App.~\ref{app:detailsOfCalSpecial} yields
\begin{eqnarray}
\label{formZ1}
Z_{{\mathcal{O}}_{{\mathcal{S}}}} = 1 + \frac{u}{6\, \epsilon}  + {\sl O} \left( u^2 \right) \ .
\end{eqnarray}

The unrenormalized theory has to be independent of the length scale $\mu^{-1}$ introduced by the renormalization. In particular the unrenormalized correlation functions with the $\mathaccent"7017{{\mathcal{O}}}_{{\mathcal{S}}}$ insertion have to be independent of $\mu$, i.e., 
\begin{eqnarray}
\label{indi}
\mu \frac{\partial}{\partial \mu} \mathaccent"7017{G}_{N,M} \left( \left\{ {\rm{\bf x}}, {\rm{\bf x}}_\parallel , \mathaccent"7017{w} \vec{\lambda}^2 \right\} ; \mathaccent"7017{u}, \mathaccent"7017{\tau}, \mathaccent"7017{c}, \mu \right)_{{\mathcal{O}}_{{\mathcal{S}}}} = 0 \ .
\end{eqnarray}
Expressed in terms of renormalized quantities, Eq.~(\ref{indi}) leads to the Gell-Mann-Low renormalization group equation
\begin{mathletters}
\label{RGequations}
\begin{eqnarray}
\left[ {\mathcal D \mathnormal}_{\mu} + \frac{N+M}{2} \gamma + \frac{M}{2} \gamma_{1} + \gamma_{{\mathcal{O}}_{{\mathcal{S}}}} \right] G_{N,M} \left( \left\{ {\rm{\bf x}}, {\rm{\bf x}}_\parallel , w \vec{\lambda}^2 \right\} ; u, \tau, c, \mu \right)_{{\mathcal{O}}_{{\mathcal{S}}}} = 0 \ ,
\end{eqnarray}
where we use the abbreviation 
\begin{eqnarray}
{\mathcal D \mathnormal}_{\mu} = \mu \frac{\partial }{\partial \mu} + \beta \frac{\partial }{\partial u} + \tau \kappa \frac{\partial }{\partial \tau} + w \zeta \frac{\partial }{\partial w} + c \kappa_c \frac{\partial }{\partial c}  \ .
\end{eqnarray}
\end{mathletters}
In Eq.~(\ref{RGequations}) we have introduced the usual Wilson functions
\begin{mathletters}
\begin{eqnarray}
\label{wilson}
\gamma_{...} \left( u \right) &=& \mu \left. \frac{\partial}{\partial \mu} \right|_{0} \ln Z_{...} \ , \\
\zeta \left( u \right) &=& \mu \left. \frac{\partial}{\partial \mu} \right|_{0} \ln w = \gamma - \gamma_w \ ,
\\
\kappa \left( u \right) &=& \mu \left. \frac{\partial}{\partial \mu} \right|_{0} \ln \tau = \gamma - \gamma_\tau\ ,
\\
\kappa_{\mathcal S \mathnormal} \left( u \right) &=& \mu \left. \frac{\partial}{\partial \mu} \right|_{0} \ln c = \gamma + \gamma_1 - \gamma_c \ ,
\\
\beta \left( u \right) &=& \mu \left. \frac{\partial u}{\partial \mu} \right|_{0} = u \left( - \epsilon + 3 \gamma - \gamma_u \right) \ ,
\end{eqnarray}
\end{mathletters}
where the bare quantities are kept fixed while taking the derivatives.

The renormalization group equation can be solved in a standard manner by the method of characteristics. The characteristics read
\begin{mathletters}
\begin{eqnarray}
l \frac{\partial \bar{\mu}}{\partial l} = \bar{\mu} \quad \mbox{with} \quad \bar{\mu}(1)=\mu \ ,
 \\
\label{char}
l \frac{\partial \bar{u}}{\partial l} = \beta \left( \bar{u}(l) \right) \quad \mbox{with} \quad \bar{u}(1)=u \ ,
 \\
l \frac{\partial}{\partial l} \ln \bar{\tau} = \kappa \left( \bar{u}(l) \right) \quad \mbox{with} \quad \bar{\tau}(1)=\tau \ ,
 \\
l \frac{\partial}{\partial l} \ln \bar{w} = \zeta \left( \bar{u}(l) \right) \quad \mbox{with} \quad \bar{w}(1)=w \ ,
 \\
l \frac{\partial}{\partial l} \ln \bar{c} = \kappa_{\mathcal{S}} \left( \bar{u}(l) \right) \quad \mbox{with} \quad \bar{c}(1)=c \ ,
 \\
l \frac{\partial}{\partial l} \ln \bar{Z}_{...} = \gamma_{...} \left( \bar{u}(l) \right) \quad \mbox{with} \quad \bar{Z}_{...}(1)=1 \ .
\end{eqnarray}
\end{mathletters}
At the infrared stable fixed point $u^\ast$, determined by $\beta \left( u^\ast \right) = 0$, we find
\begin{eqnarray}
\label{solRG}
\lefteqn{ G_{N,M} \left( \left\{ {\rm{\bf x}}, {\rm{\bf x}}_\parallel , w \vec{\lambda}^2 \right\} ; u, \tau, c, \mu \right)_{{\mathcal{O}}_{{\mathcal{S}}}} = l^{(N+M)\eta /2 + M\eta_{1}/2 + \gamma^\ast_{{\mathcal{O}}_{{\mathcal{S}}}}} }
\nonumber \\ && \times
G_{N,M} \left( \left\{ l {\rm{\bf x}}, l {\rm{\bf x}}_\parallel , l^{\zeta^\ast} w \vec{\lambda}^2 \right\} ; u^\ast , l^{\kappa^\ast} \tau, l^{\kappa^\ast_{{\mathcal{S}}}} c, \mu \right)_{{\mathcal{O}}_{{\mathcal{S}}}} \ ,
\end{eqnarray}
where $\gamma^\ast = \gamma \left( u^\ast \right)$, $\gamma^\ast_1 = \gamma_1 \left( u^\ast \right)$, $\kappa^\ast = \kappa \left( u^\ast \right)$, $\kappa^\ast_{{\mathcal{S}}} = \kappa_{{\mathcal{S}}} \left( u^\ast \right)$ and $\gamma^\ast_{{\mathcal{O}}_{{\mathcal{S}}}} = \gamma_{{\mathcal{O}}_{{\mathcal{S}}}} \left( u^\ast \right)$. In order to obtain the scaling behavior of the correlation functions, a dimensional analysis remains to be done. This dimensional analysis gives
\begin{eqnarray}
\label{dimA}
\lefteqn{ G_{N,M} \left( \left\{ {\rm{\bf x}}, {\rm{\bf x}}_\parallel , w \vec{\lambda}^2 \right\} ; u, \tau, c, \mu \right)_{{\mathcal{O}}_{{\mathcal{S}}}} = \mu^{(N+M)(d-2) /2 + 1} }
\nonumber \\
&& \times
G_{N,M} \left( \left\{ \mu {\rm{\bf x}}, \mu {\rm{\bf x}}_\parallel , \mu^{-2} w \vec{\lambda}^2 \right\} ; u^\ast , \mu^{-2} \tau, \mu^{-1} c, \mu \right)_{{\mathcal{O}}_{{\mathcal{S}}}} \ .
\end{eqnarray}
Equation~(\ref{solRG}) in conjunction with Eq.~(\ref{dimA}) now results in
\begin{eqnarray}
\label{skalenverhalten}
\lefteqn{ G_{N,M} \left( \left\{ {\rm{\bf x}}, {\rm{\bf x}}_\parallel , w \vec{\lambda}^2 \right\} ; u, \tau, c, \mu \right)_{{\mathcal{O}}_{{\mathcal{S}}}} = l^{(N+M)(d-2+\eta) /2 + M\eta_{1}/2 + 1 + \gamma^\ast_{{\mathcal{O}}_{{\mathcal{S}}}}} }
\nonumber \\ && \times
G_{N,M} \left( \left\{ l {\rm{\bf x}}, l {\rm{\bf x}}_\parallel , l^{-\phi /\nu} w \vec{\lambda}^2 \right\} ; u^\ast , l^{-1/\nu} \tau, l^{-1/\nu_{{\mathcal{S}}}} c, \mu \right)_{{\mathcal{O}}_{{\mathcal{S}}}} \ ,
\end{eqnarray}
where $\eta = \gamma^\ast$, $1/\nu = 2 - \kappa^\ast$, $\eta_{1} = \gamma_1^\ast$, and $1/\nu_{\mathcal S \mathnormal} = 1 - \kappa_{\mathcal S \mathnormal}^\ast$ are usual critical exponents for semi-infinite percolation, whose $\epsilon$ expansion results may be inferred, e.g., from\cite{alcantara_80,diehl_lam_89}. $\phi = \nu \left( 2 - \zeta^\ast \right) = 1+\epsilon /42 + 4\epsilon^2 / 3087 + {\sl O} \left( \epsilon^3 \right)$ is the bulk resistance exponent\cite{stenull_janssen_oerding_99}. 

The scaling behavior of the correlation functions without insertion can be derived by similar means. The corresponding renormalization group equation is analogous to Eq.~(\ref{RGequations}). Basically, just the $\gamma_{{\mathcal{O}}_{{\mathcal{S}}}}$ is missing. Solving this renormalization group equation gives in conjunction with dimensional analysis
\begin{eqnarray}
\label{skalenverhaltenPur}
\lefteqn{ G_{N,M} \left( \left\{ {\rm{\bf x}}, {\rm{\bf x}}_\parallel , w \vec{\lambda}^2 \right\} ; u, \tau, c, \mu \right) = l^{(N+M)(d-2+\eta) /2 + M\eta_{1}/2 } }
\nonumber \\ && \times
G_{N,M} \left( \left\{ l {\rm{\bf x}}, l {\rm{\bf x}}_\parallel , l^{-\phi /\nu} w \vec{\lambda}^2 \right\} ; u^\ast , l^{-1/\nu} \tau, l^{-1/\nu_{{\mathcal{S}}}} c, \mu \right) \ .
\end{eqnarray}
By comparing Eqs.~(\ref{skalenverhalten}) and (\ref{skalenverhaltenPur}) we learn that the scaling dimension $x_{{\mathcal{O}}_{{\mathcal{S}}}}$ of ${\mathcal{O}}_{{\mathcal{S}}}$ is
\begin{eqnarray}
\label{skalenDim}
x_{{\mathcal{O}}_{{\mathcal{S}}}}  = 1 + \gamma_{{\mathcal{O}}_{{\mathcal{S}}}}^\ast \ ,
\end{eqnarray}
which tells us that the coupling $v_{{\mathcal{S}}}$ scales in terms of the flow parameter $l$ as
\begin{eqnarray}
v_{{\mathcal{S}}} (l) = v_{{\mathcal{S}}} \, l^{1 + \gamma_{{\mathcal{O}}_{{\mathcal{S}}}}^\ast} \ .
\end{eqnarray}
Taking into account that
\begin{eqnarray}
w (l) =w \, l^{\zeta^\ast -2} 
\end{eqnarray}
we deduce that
\begin{eqnarray}
w_{{\mathcal{S}}} (l) = w_{{\mathcal{S}}} \, l^{- \phi_{\mathcal{S}}/\nu } \ .
\end{eqnarray}
Here we introduced the surface resistance exponent $\phi_{\mathcal{S}}$ for the special transition. To one-loop order it is given by
\begin{eqnarray}
\phi_{\mathcal{S}} = \nu \left(  1 - \zeta^\ast - \gamma_{{\mathcal{O}}_{{\mathcal{S}}}}^\ast \right)= \frac{1}{2} - \frac{1}{84} \, \epsilon + {\sl O} \left( \epsilon^2 \right) \ .
\end{eqnarray}

Now we can extract the scaling behavior of the average resistance between surface ports from the correlation functions with $N=0$ and $M=2$. At the special transition, i.e., for $\tau = c =0$, Eqs.~(\ref{skalenverhalten}) and (\ref{skalenverhaltenPur}) reduce to
\begin{eqnarray}
\label{skalenverhaltenZweiPunkt}
G_{0,2} \left( \left| {\rm{\bf x}}_\parallel - {\rm{\bf x}}_\parallel^\prime \right|  , w \vec{\lambda}^2 \right)_{{\mathcal{O}}_{{\mathcal{S}}}} = l^{(d-2+\eta_\parallel) - ( \phi_{\mathcal{S}} - \phi )/\nu} 
G_{0,2} \left( l \left| {\rm{\bf x}}_\parallel - {\rm{\bf x}}_\parallel^\prime \right| , l^{-\phi /\nu} w \vec{\lambda}^2 \right)_{{\mathcal{O}}_{{\mathcal{S}}}} 
\end{eqnarray}
and, respectively,
\begin{eqnarray}
\label{skalenverhaltenZweiPunktPur}
G_{0,2} \left( \left| {\rm{\bf x}}_\parallel - {\rm{\bf x}}_\parallel^\prime \right|  , w \vec{\lambda}^2 \right) = l^{(d-2+\eta_\parallel)} 
G_{0,2} \left( l \left| {\rm{\bf x}}_\parallel - {\rm{\bf x}}_\parallel^\prime \right| , l^{-\phi /\nu} w \vec{\lambda}^2 \right) \ ,
\end{eqnarray}
where $\eta_\parallel = \eta + \eta_1$. Note that we have dropped some of the arguments for notational simplicity. Clearly, the two-point function with and the one without the insertion enter into the average resistance. Overall, we may write for the generating function
\begin{eqnarray}
\label{overall}
G \left( \left| {\rm{\bf x}}_\parallel - {\rm{\bf x}}_\parallel^\prime \right|  , w \vec{\lambda}^2 \right) &=& l^{(d-2+\eta_\parallel)} \bigg\{
G_{0,2} \left( l \left| {\rm{\bf x}}_\parallel - {\rm{\bf x}}_\parallel^\prime \right| , l^{-\phi /\nu} w \vec{\lambda}^2 \right)
\nonumber \\ 
&& + \,  l^{- ( \phi_{\mathcal{S}} - \phi )/\nu} G_{0,2} \left( l \left| {\rm{\bf x}}_\parallel - {\rm{\bf x}}_\parallel^\prime \right| , l^{-\phi /\nu} w \vec{\lambda}^2 \right)_{{\mathcal{O}}_{{\mathcal{S}}}} \bigg\} \ .
\end{eqnarray}
We exploit the freedom to choose $l = \left| {\rm{\bf x}}_\parallel - {\rm{\bf x}}_\parallel^\prime \right|^{-1}$. Moreover, we carry out a Taylor expansion which yields
\begin{eqnarray}
\label{overall2}
G \left( \left| {\rm{\bf x}}_\parallel - {\rm{\bf x}}_\parallel^\prime \right|  , w \vec{\lambda}^2 \right) &=& \left| {\rm{\bf x}}_\parallel - {\rm{\bf x}}_\parallel^\prime \right|^{(d-2+\eta_\parallel)} \bigg\{ 1 + \left| {\rm{\bf x}}_\parallel - {\rm{\bf x}}_\parallel^\prime \right|^{\phi /\nu} w \vec{\lambda}^2 
\nonumber \\ 
&& + \,  \frac{w_{\mathcal{S}}}{w} \left| {\rm{\bf x}}_\parallel - {\rm{\bf x}}_\parallel^\prime \right|^{( \phi_{\mathcal{S}} - \phi )/\nu} \left| {\rm{\bf x}}_\parallel - {\rm{\bf x}}_\parallel^\prime \right|^{\phi /\nu} w \vec{\lambda}^2 + \cdots \bigg\} \ .
\end{eqnarray}
Here we have set all non universal constants equal to one. Upon differentiating Eq.~(\ref{overall2}) with respect to $\vec{\lambda}^2$ we obtain at $\vec{\lambda}^2 = 0$ for the average resistance
\begin{eqnarray}
\label{aveRes}
M_R \left( {\rm{\bf x}}_\parallel , {\rm{\bf x}}_\parallel^\prime \right)  = w \left| {\rm{\bf x}}_\parallel - {\rm{\bf x}}_\parallel^\prime \right|^{\phi /\nu} \bigg\{ 1 +  \frac{w_{\mathcal{S}}}{w} \left| {\rm{\bf x}}_\parallel - {\rm{\bf x}}_\parallel^\prime \right|^{( \phi_{\mathcal{S}} - \phi )/\nu} + \cdots \bigg\} \ .
\end{eqnarray}
Note that $( \phi_{\mathcal{S}} - \phi )/\nu < 0$, i.e., the corresponding term is indeed only a correction to scaling that vanishes for $\left| {\rm{\bf x}}_\parallel - {\rm{\bf x}}_\parallel^\prime \right| \to \infty$.

\subsection{Ordinary transition}
At the ordinary transition the order parameter field satisfies the Dirichlet boundary condition $\varphi_{\mathcal S \mathnormal} = 0$. This can be deduced from Eq.~(\ref{semiInfProp}), which shows that the Gaussian propagator vanishes for $c\to \infty$ if one of the points is located on the surface. Since correlation functions with insertions of the surface field are zero, they are not appropriate to gain information about fluctuation effects near the surface. A convenient method to investigate the scaling behavior of quantities that vanish for $c\to \infty$ is the $1/c$ expansion\cite{diehl_dietrich_80,diehl_dietrich_eisenriegler_83}. For large $c$ the propagator behaves as
\begin{eqnarray}
G_{{\rm{\bf p}},\vec{\lambda}} \left( z, 0 \right) = c^{-1} \partial_{z^\prime} G^{D}_{{\rm{\bf p}},\vec{\lambda}} \left( z, z^\prime \right) \Big|_{z^\prime =0} + {\sl O} \left( c^{-2} \right) \ ,
\end{eqnarray}
where $G^{D}_{{\rm{\bf p}},\vec{\lambda}} \left( z, z^\prime \right) = \lim_{c\to \infty} G_{{\rm{\bf p}},\vec{\lambda}} \left( z, z^\prime \right)$ is the Dirichlet propagator. Similarly one can exand the entire correlation functions which gives
\begin{eqnarray}
\label{corrFktExp}
G_{N,M} \left( \left\{ {\rm{\bf x}}, {\rm{\bf x}}_{\|}, \vec{\theta} \right\} \right) = c^{-1} \delta_{N,0} \, \delta_{M,2} \,\delta \left( {\rm{\bf x}}_{\|} - {\rm{\bf x}}_{\|}^\prime \right) + c^{-M} G^{\infty}_{N,M} \left( \left\{ {\rm{\bf x}}, {\rm{\bf x}}_{\|}, \vec{\theta} \right\} \right) + \cdots
\end{eqnarray}
where 
\begin{eqnarray}
G^{\infty}_{N,M} \left( \left\{ {\rm{\bf x}}, {\rm{\bf x}}_{\|}, \vec{\theta} \right\} \right) = \lim_{D \to 0} \left\langle \prod_{i=1}^{N} \varphi \left( {\rm{\bf x}}_{i} , \vec{\theta}_{i} \right) \prod_{j=1}^{M} \partial_n \varphi_{\mathcal S \mathnormal} \left( {\rm{\bf x}}_{\| j} , \vec{\theta}_{j} \right) \right\rangle^{\mbox{\scriptsize{conn}}}_{{\mathcal H \mathnormal}, \infty} \ .
\end{eqnarray}
The subscript $\infty$ on the right hand side reminds us that the average is to be taken with $c =\infty$. Apart from the additional term for $N=0$, $M=2$ the $1/c$ expansion amounts in replacing all surface fields $\varphi_{\mathcal S \mathnormal}$ by $\partial_n \varphi_{\mathcal S \mathnormal}$. The surface operator $\partial_n \varphi_{\mathcal S \mathnormal}$ may be renormalized by the reparametrization
\begin{eqnarray}
\partial_n \varphi_{\mathcal S \mathnormal} \to \left[ \partial_n \varphi_{\mathcal S \mathnormal} \right]_{\mbox{\scriptsize{bare}}} = \left( Z_{1}^{\infty} Z \right)^{1/2} \partial_n \varphi_{\mathcal S \mathnormal} \ .
\end{eqnarray}
A one-loop result for $Z_{1}^{\infty}$ can be gleaned from\cite{diehl_lam_89}.

We point out that Eq.~(\ref{corrFktExp}) is correct in the given form for the bare, i.e., unrenormalized correlation functions. In the renormalized theory nontrivial powers of $1/c$ will appear instead of $1/c$. One may inferre from Eq.~(\ref{ableitungsvorschrift}), however, that this peculiarity will not influence the average resistance.

Here at the ordinary transition we insert the operator
\begin{eqnarray}
{\mathcal O \mathnormal}^\infty_{\mathcal S \mathnormal} =  \frac{w}{2} \int d^{d-1} x_\parallel \, \sum_{\vec{\theta}} \, \left( \nabla_{\vec{\theta}} \, \partial_n \varphi_{\mathcal S \mathnormal} \left( {\rm{\bf x}}_{\|} , \vec{\theta} \right) \right)^2
\end{eqnarray}
into the one-loop diagrams contributing to $G^{\infty}_{2,0}$. The corresponding diagrams are depicted in Fig.~\ref{ordinaryDiagrams}. We use the renormalization
\begin{eqnarray}
{\mathcal{O}}^\infty_{{\mathcal{S}}} \to \mathaccent"7017{{\mathcal{O}}}^\infty_{{\mathcal{S}}} &=& Z_{{\mathcal{O}}^\infty_{{\mathcal{S}}}} {\mathcal{O}}^\infty_{{\mathcal{S}}} \ .
\end{eqnarray}
Our one-loop calculation gives for the renormalization factor
\begin{eqnarray}
Z_{\mathcal{O}^\infty_{\mathcal S \mathnormal}} = 1 + \frac{23}{30} \frac{u}{\epsilon} + {\sl O} \left( u^2 \right) \ .
\end{eqnarray}
Details of the calculation can be found in appendix~\ref{app:detailsOfCalOrdinary}.

The renormalization group equation for the correlation functions $G^{\infty}_{N,M}$ with the $\mathcal{O}^\infty_{\mathcal S \mathnormal}$ insertion is given by
\begin{mathletters}
\label{RGequationsOrdnary}
\begin{eqnarray}
\left[ {\mathcal D \mathnormal}^{\infty}_{\mu} + \frac{N+M}{2} \gamma + \frac{M}{2} \gamma_{1}^\infty + \gamma_{{\mathcal{O}}^\infty_{{\mathcal{S}}}} \right] G^\infty_{N,M} \left( \left\{ {\rm{\bf x}}, {\rm{\bf x}}_\parallel , w \vec{\lambda}^2 \right\} ; u, \tau, \mu \right)_{{\mathcal{O}}^\infty_{{\mathcal{S}}}} = 0 \ ,
\end{eqnarray}
where ${\mathcal D \mathnormal}^\infty_{\mu}$ is an abbreviation for
\begin{eqnarray}
{\mathcal D \mathnormal}^\infty_{\mu} = \mu \frac{\partial }{\partial \mu} + \beta \frac{\partial }{\partial u} + \tau \kappa \frac{\partial }{\partial \tau} + w \zeta \frac{\partial }{\partial w}  \ .
\end{eqnarray}
\end{mathletters}
The Wilson functions $\gamma_{1}^\infty$ and $\gamma_{{\mathcal{O}}^\infty_{{\mathcal{S}}}}$ are defined as
\begin{eqnarray}
\gamma^\infty_{1} \left( u \right) &=& \mu \left. \frac{\partial}{\partial \mu} \right|_{0} \ln Z^\infty_{1} \,
\nonumber \\
\gamma_{{\mathcal{O}}^\infty_{{\mathcal{S}}}} \left( u \right) &=& \mu \left. \frac{\partial}{\partial \mu} \right|_{0} \ln Z_{\mathcal{O}^\infty_{\mathcal S \mathnormal}} \ .
\end{eqnarray}
Solving Eq.~(\ref{RGequationsOrdnary}) provides us in conjunction with a dimensional analysis with the scaling behavior
\begin{eqnarray}
\label{skalenverhaltenOrdinary}
\lefteqn{ G_{N,M}^\infty \left( \left\{ {\rm{\bf x}}, {\rm{\bf x}}_\parallel , w \vec{\lambda}^2 \right\} ; u, \tau, \mu \right)_{{\mathcal{O}}_{{\mathcal{S}}}} = l^{(N+M)(d-2+\eta) /2 + M\eta^\infty_{1}/2 + 3 + \gamma^\ast_{{\mathcal{O}}^\infty_{{\mathcal{S}}}}} }
\nonumber \\ && \times
G_{N,M}^\infty \left( \left\{ l {\rm{\bf x}}, l {\rm{\bf x}}_\parallel , l^{-\phi /\nu} w \vec{\lambda}^2 \right\} ; u^\ast , l^{-1/\nu} \tau, \mu \right)_{{\mathcal{O}}^\infty_{{\mathcal{S}}}} \ ,
\end{eqnarray}
where $\eta^\infty_1 = \gamma^\infty_{1} \left( u^\ast \right)$ and $\gamma_{{\mathcal{O}}^\infty_{{\mathcal{S}}}}^\ast = \gamma_{{\mathcal{O}}^\infty_{{\mathcal{S}}}} \left( u^\ast \right)$. For the correlation functions $G^{\infty}_{N,M}$ without insertion we obtain
\begin{eqnarray}
\label{skalenOrdinaryPur}
\lefteqn{ G_{N,M}^\infty \left( \left\{ {\rm{\bf x}}, {\rm{\bf x}}_\parallel , w \vec{\lambda}^2 \right\} ; u, \tau, \mu \right) = l^{(N+M)(d-2+\eta) /2 + M\eta^\infty_{1}/2} }
\nonumber \\ && \times
G_{N,M}^\infty \left( \left\{ l {\rm{\bf x}}, l {\rm{\bf x}}_\parallel , l^{-\phi /\nu} w \vec{\lambda}^2 \right\} ; u^\ast , l^{-1/\nu} \tau, \mu \right) \ .
\end{eqnarray}
By comparing Eqs.~(\ref{skalenverhaltenOrdinary}) and (\ref{skalenOrdinaryPur}) we deduce that the coupling constant $v_{\mathcal{S}}^\infty = w_{\mathcal{S}} / \left( c^2 \, w \right)$ of ${\mathcal{O}^\infty_{\mathcal S \mathnormal}}$ scales as
\begin{eqnarray}
v_{\mathcal{S}}^\infty (l) = v_{\mathcal{S}}^\infty \, l^{3 + \gamma^\ast_{{\mathcal{O}}^\infty_{{\mathcal{S}}}}} \ .
\end{eqnarray}
Hence we obtain for $w_{\mathcal{S}}^\infty = w_{\mathcal{S}} / c^2$ that
\begin{eqnarray}
w_{\mathcal{S}}^\infty (l) = w_{\mathcal{S}}^\infty \, l^{-\phi_{\mathcal{S}}^\infty/\nu} \ ,
\end{eqnarray}
with the surface resistance exponent $\phi_{\mathcal{S}}^\infty$ for the ordinary transition reading
\begin{eqnarray}
\phi_{\mathcal{S}}^\infty = \nu \left( - 1 - \zeta^\ast - \gamma_{{\mathcal{O}}^\infty_{{\mathcal{S}}}}^\ast \right)= - \frac{1}{2} - \frac{19}{420} \, \epsilon + {\sl O} \left( \epsilon^2 \right) \ .
\end{eqnarray}

Now we can proceed in the same fashion as in Sec.~\ref{specialTrans}. We assemble the generating function for the average resistance from the two-point correlation functions ($N=0$, $M=2$) with and without insertion. Upon taking the derivative with respect to $\vec{\lambda}^2$ we finally arrive at
\begin{eqnarray}
\label{aveResOrdinary}
M_R \left( {\rm{\bf x}}_\parallel , {\rm{\bf x}}_\parallel^\prime \right)  = w \left| {\rm{\bf x}}_\parallel - {\rm{\bf x}}_\parallel^\prime \right|^{\phi /\nu} \bigg\{ 1 +  \frac{w_{\mathcal{S}}^\infty}{w} \left| {\rm{\bf x}}_\parallel - {\rm{\bf x}}_\parallel^\prime \right|^{( \phi_{\mathcal{S}}^\infty - \phi )/\nu} + \cdots \bigg\} \ .
\end{eqnarray}

\subsection{Leading correction}
Now, as we have computed the corrections to scaling due the resistors located on the surface it is  legitimate to ask: How important are these corrections compared to other corrections? In particular we should compare them to the leading correction which is governed by the so called Wegner exponent. This leading correction emerges when the renormalized coupling $u$ is not exactly equal $u^\ast$ since the renormalization flow has not arrived at its fixed point yet. Such a case occurs typically when there is a finite momentum cutoff reminiscent of a nonvanishing lattice spacing.

In order to determine the leading correction we revisit the characteristic Eq.~(\ref{char}). Upon expansion for small deviations $u - u^\ast$ from $u^\ast$ we obtain
\begin{eqnarray}
l \frac{\partial \bar{u}}{\partial l} = \omega \, \left[ \bar{u} - u^\ast \right] + {\sl O} \left(\left[ u - u^\ast \right]^2  \right) \quad \mbox{with} \quad \bar{u}(1)=u \ ,
\end{eqnarray}
where $\omega = \beta^\prime \left( u^\ast \right)$. This differential equation is readily solved with the result
\begin{eqnarray}
\label{charDev}
\bar{u} (l) =  u^\ast + \left[ u - u^\ast \right] l^\omega \ .
\end{eqnarray}
$\omega$ is referred to as the Wegner exponent. It can be calculated without much effort to third order in $\epsilon$ upon using the three-loop result for $\beta \left( u \right)$ obtained by de Alcantara Bonfim {\it et al}.~\cite{alcantara_80}. Here we are working only to first order in $\epsilon$ to which the Wegner exponent is given by
\begin{eqnarray}
\omega = \epsilon + {\sl O} \left( \epsilon^2  \right) \ .
\end{eqnarray}

To see the impact of the deviation from $u^\ast$ on the correlation functions, we revisit the special transition. The analysis can be adapted, however, in an obvious fashion to the ordinary transition. From the renormalization group equation and Eq.~(\ref{charDev}) we deduce that
\begin{eqnarray}
\label{skalenverhaltenWegner}
G_{0,2} \left( \left| {\rm{\bf x}}_\parallel - {\rm{\bf x}}_\parallel^\prime \right|  , w \vec{\lambda}^2 \right) &=& l^{(d-2+\eta_\parallel)} 
G_{0,2} \left( l \left| {\rm{\bf x}}_\parallel - {\rm{\bf x}}_\parallel^\prime \right| , l^{-\phi /\nu} w \vec{\lambda}^2 \right)
\nonumber \\
&\times& \bigg\{ 1 + \left[ u - u^\ast \right] l^\omega  F  \left( l \left| {\rm{\bf x}}_\parallel - {\rm{\bf x}}_\parallel^\prime \right| , l^{-\phi /\nu} w \vec{\lambda}^2 \right) + {\sl O} \left(\left[ u - u^\ast \right]^2  \right) \bigg\} \ ,
\end{eqnarray}
where $F$ is a scaling function. Once more we choose $l = \left| {\rm{\bf x}}_\parallel - {\rm{\bf x}}_\parallel^\prime \right|^{-1}$ and carry out a Taylor expansion,
\begin{eqnarray}
\label{WegnerExpanded}
G_{0,2} \left( \left| {\rm{\bf x}}_\parallel - {\rm{\bf x}}_\parallel^\prime \right|  , w \vec{\lambda}^2 \right) &=& \left| {\rm{\bf x}}_\parallel - {\rm{\bf x}}_\parallel^\prime \right|^{(d-2+\eta_\parallel)} 
\bigg[ 1+ \left| {\rm{\bf x}}_\parallel - {\rm{\bf x}}_\parallel^\prime \right|^{\phi /\nu} w \vec{\lambda}^2 + \cdots \bigg]
\nonumber \\
&\times& \bigg\{ 1 + \left[ u - u^\ast \right] \left| {\rm{\bf x}}_\parallel - {\rm{\bf x}}_\parallel^\prime \right|^{-\omega} \bigg[ 1 + \left| {\rm{\bf x}}_\parallel - {\rm{\bf x}}_\parallel^\prime \right|^{\phi /\nu} w \vec{\lambda}^2 + \cdots \bigg] + \cdots  \bigg\} \ ,
\end{eqnarray}
where we have set, as always, all non universal expansion coefficients equal to one. By taking the derivative with respect to $\vec{\lambda}^2$ we obtain at $\vec{\lambda}^2 = 0$
\begin{eqnarray}
\label{aveResWegner}
M_R \left( {\rm{\bf x}}_\parallel , {\rm{\bf x}}_\parallel^\prime \right)  = w \left| {\rm{\bf x}}_\parallel - {\rm{\bf x}}_\parallel^\prime \right|^{\phi /\nu} \bigg\{ 1 + \left[ u - u^\ast \right]  \left| {\rm{\bf x}}_\parallel - {\rm{\bf x}}_\parallel^\prime \right|^{-\omega} + \cdots \bigg\} \ .
\end{eqnarray}

Now we can compare the different corrections to $M_R$ considered in this paper. To first order in $\epsilon$ the correction due to the deviation from $u^\ast$ falls off algebraically for increasing port separation with an exponent $-\omega = -\epsilon$. The surface correction at the special transition vanishes much faster with an exponent $(\phi_{\mathcal{S}} - \phi )/\nu = -1 + \epsilon/21$. Among the three corrections, the surface correction at the ordinary transition drops off fastest with $(\phi_{\mathcal{S}}^\infty - \phi )/\nu = -3 + 23 \epsilon/105$.

\section{Conclusions}
\label{Conclusions} 
In order to study the effects of surfaces on resistor percolation we have considered a semi-infinte RRN. We have presented a field theoretic Hamiltonian in which the coupling constant corresponding to the surface conductances turned out to be irrelevant. We have calculated the corrections to scaling due to this irrelevant coupling for the special and the ordinary transition to one-loop order.

In this paper we did not consider the surface and the extraordinary transition. We left the surface transition aside, because it is basically equivalent to the percolation transition of a translationally invariant $d-1$ dimensional RRN. Thus, the behavior of $M_R$ at the surface transition can be inferred from Ref.~\cite{stenull_janssen_oerding_99}. We did not drill into the extraordinary transition because of severe technical complications. These are rooted in the fact that the order parameter profile is not flat on neither side of the line of the extraordinary transition. 

That the surface coupling $w_{\mathcal{S}}$ is irrelevant seems intuitively plausible. Suppose that the resistor network is at the special transition where the surface and the bulk percolate simultaneously. Assume that we conduct a series of consecutive measurements in which we apply the external current $I$ between two surface ports $x_\parallel$ and $x_\parallel^\prime$ with increasing distance $\left| x_\parallel - x_\parallel^\prime \right|$. As we increase $\left| x_\parallel - x_\parallel^\prime \right|$ more and more paths of connected bulk resistors will add in parallel to the connected paths of surface resistors. Hence, the influence of the surface bonds becomes negligible for large $\left| x_\parallel - x_\parallel^\prime \right|$. This situation is even more pronounced at the ordinary transition, where the percolation probability $P_{\mbox{\scriptsize{perc}}}$ is lower at the surface than in the bulk. Thus, it is plausible that the surface correction to $M_R$ vanishes faster for increasing $\left| x_\parallel - x_\parallel^\prime \right|$ at the ordinary transition than at the special transition.

Both surface corrections turn out to be small compared to the leading correction governed by the Wegner exponent for percolation. This means that the surface has weak effects on the average resistance compared to those of a finite lattice spacing. As long as one is interested only in the leading behavior and mayor corrections to it, one may neglect the  surface effects safely.

To our knowledge there are no numerical simulations available to date which could be used to test our predictions for the surface resistance exponents. The reason is probably that the typical simulations reported in the literature use a so called bus-bar geometry, in which the resistor network is placed between superconducting plates which short entire surfaces. We hope that this paper triggers simulations providing numerical estimates for $\phi_{\mathcal{S}}$ and $\phi_{\mathcal{S}}^\infty$.

\acknowledgements
We acknowledge support by the Sonderforschungsbereich 237 ``Unordnung und gro{\ss}e Fluktuationen'' of the Deutsche Forschungsgemeinschaft. 

\appendix
\section{Evaluation of Diagrams for the special transition}
\label{app:detailsOfCalSpecial}
In this appendix we sketch the computation of $Z_{\mathcal{O}_{\mathcal S \mathnormal}}$ for the special transition. We start with diagram a and revisit Eq.~(\ref{surfaceDiagram_a}). Upon inserting the result for the current summation, Eq.~(\ref{currentSum3}), we obtain \begin{eqnarray}
\mbox{a} &=& - g^2 w \vec{\lambda}^2 \frac{1}{\left( 4 \pi \right)^{3/2}} \int_{{\rm{\bf p}}} \int_0^\infty \frac{ds_1 ds_2 ds_3}{\sqrt{s_1 s_2 s_3}} \left( \frac{s_1}{s_1 + s_2 + s_3} \right)^2 \exp \left[ - \left( s_1 + s_2 + s_3 \right) \left( \tau + {\rm{\bf p}}^2 \right) \right] 
\nonumber \\
&\times& \exp \left[ - \frac{z^2}{4s_2} - \frac{z^{\prime 2}}{4s_3} \right] \left\{ \exp \left[ - \frac{\left( z - z^\prime \right)^2}{4s_1} \right] + \exp \left[ - \frac{\left( z + z^\prime \right)^2}{4s_1} \right] \right\} \ ,
\end{eqnarray}
where we have dropped all other terms since we are interested here only in  the part of a proportional to $w$. The momentum integration is straightforward and yields
\begin{eqnarray}
\mbox{a} &=& - g^2 w \vec{\lambda}^2 \frac{1}{\left( 4 \pi \right)^{3/2}} \int_0^\infty \frac{ds_1 ds_2 ds_3}{\sqrt{s_1 s_2 s_3}} \left( \frac{s_1}{s_1 + s_2 + s_3} \right)^2 \left( \frac{1}{4 \pi \left( s_1 + s_2 + s_3 \right)} \right)^{(d-2)/2}  
\nonumber \\
&\times& \exp \left[ - \left( s_1 + s_2 + s_3 \right) \tau  \right] \exp \left[ - \frac{z^2}{4s_2} - \frac{z^{\prime 2}}{4s_3} \right] \left\{ \exp \left[ - \frac{\left( z - z^\prime \right)^2}{4s_1} \right] + \exp \left[ - \frac{\left( z + z^\prime \right)^2}{4s_1} \right] \right\} \ .
\end{eqnarray}
At this stage it is useful to apply the Laplace transformation. The Laplace transformed of a reads
\begin{eqnarray}
\label{guckuck}
{\mathcal L \mathnormal} \left( \mbox{a} \right) &=& \int_0^\infty dz \int_0^\infty dz^\prime \exp \left[ - uz -vz^\prime \right] \mbox{a}
\nonumber \\
&=& - g^2 w \vec{\lambda}^2 \frac{2}{\left( 4 \pi \right)^{d/2}} \int_0^\infty ds_1 ds_2 ds_3 \ \exp \left[ - \left( s_1 + s_2 + s_3 \right) \tau  \right] 
\nonumber \\
&\times& \frac{s_1^2}{\left( s_1 + s_2 + s_3 \right)^{2+d/2}} + {\sl O} \left( u,v \right) \ ,
\end{eqnarray}
where we dropped higher order terms in $u$ and $v$ since they are convergent. The integrations in Eq.~(\ref{guckuck}) are simplified by the change of variables $s_1 \to tx$, $s_2 \to ty$, and $s_3 \to t \left( 1 - x - y \right)$:
\begin{eqnarray}
{\mathcal L \mathnormal} \left( \mbox{a} \right) &=&  - g^2 w \vec{\lambda}^2 \frac{2}{\left( 4 \pi \right)^{d/2}} \int_0^1 dx \int_0^{1-x} dy \int_0^\infty dt \ x^2 t^{2-d/2} \exp \left[ - t \tau  \right] \ .
\end{eqnarray}
The result
\begin{eqnarray}
{\mathcal L \mathnormal} \left( \mbox{a} \right) &=&  - g^2 w \vec{\lambda}^2 \frac{1}{\left( 4 \pi \right)^{d/2}} \frac{1}{6} \tau^{d/2-3} \Gamma \left( 3 - \frac{d}{2} \right)
\end{eqnarray}
can be expanded for small $\epsilon = 6-d$ as
\begin{eqnarray}
{\mathcal L \mathnormal} \left( \mbox{a} \right) &=&  - g^2 w \vec{\lambda}^2 \tau^{-\epsilon/2} \frac{G_\epsilon}{3\epsilon} \ .
\end{eqnarray}
Upon transforming back to real space we obtain
\begin{eqnarray}
\label{realSpaceResult_a}
\mbox{a} = - g^2 w \vec{\lambda}^2  \frac{G_\epsilon}{3\epsilon} \tau^{-\epsilon /2} \delta_+ \left( z^\prime \right) \delta_+ \left( z^{\prime \prime} \right) \ ,
\end{eqnarray}
with $\delta_+ \left( z \right)$ denoting the distribution defined by
\begin{eqnarray}
\label{a10}
\int_0^\infty \delta_+ \left( z \right) g \left( z \right) = g \left( 0 \right) \ .
\end{eqnarray}

Alternatively, a can be computed with help of the parameter sum
\begin{eqnarray}
\label{parameterSum}
\Sigma \left( n, m, k \right)  = \lim_{D \to 0} \sum_{\vec{\lambda}^{(l)}} \vec{\lambda}^{(l)2} \frac{1}{\kappa_{\vec{\lambda}+\vec{\lambda}^{(l)}}^n \kappa_{\vec{\lambda}}^m \left( \kappa_{\vec{\lambda}+\vec{\lambda}^{(l)}} + \kappa_{\vec{\lambda}} \right)^k} \ .
\end{eqnarray}
The evaluation of this sum yielding
\begin{eqnarray}
\label{resParameterSum}
\Sigma \left( n, m, k \right)  = \vec{\lambda}^{2} \frac{1}{\kappa_{\vec{0}}^{n+m+k}} \, 2^{-k} \, \frac{ n \left( n+2 \right) + \left( n+1 \right) k + \frac{1}{4} k \left( k-1 \right) }{ \left( n+m+k \right) \left( n+m+k+2 \right) } + {\sl O} \left( \left( \vec{\lambda}^{2}\right)^2 \right)
\end{eqnarray}
is outlined in appendix~\ref{app:evalParamSum}. In this approach we do not parametrize the Neumann propagator. Instead, we substitute Eq.~(\ref{NeumannProp}) directly into Eq.~(\ref{surfaceDiagram_a}). Laplace transformation then gives
\begin{eqnarray}
\label{wolf}
{\mathcal L \mathnormal} \left( \mbox{a} \right) &=& - g^2 w \sum_{\vec{\lambda}^{(l)}} \int_{{\rm{\bf p}}} \vec{\lambda}^{(l)2} \frac{1}{2 \kappa_{\vec{\lambda}+\vec{\lambda}^{(l)}} \kappa_{\vec{\lambda}}^2}
\left[ \frac{1}{2\kappa_{\vec{\lambda}} + u + v} \left( \frac{1}{\kappa_{\vec{\lambda}+\vec{\lambda}^{(l)}} + \kappa_{\vec{\lambda}} + u} + \frac{1}{\kappa_{\vec{\lambda}+\vec{\lambda}^{(l)}} + \kappa_{\vec{\lambda}} + v} \right)
\right.
\nonumber \\ 
&+& \left. \frac{1}{\kappa_{\vec{\lambda}+\vec{\lambda}^{(l)}} + \kappa_{\vec{\lambda}} + u} \, \frac{1}{\kappa_{\vec{\lambda}+\vec{\lambda}^{(l)}} + \kappa_{\vec{\lambda}} + v} \right] \ .
\end{eqnarray}
Taylor expansion of the right hand side of Eq.~(\ref{wolf}) leads to
\begin{eqnarray}
\label{wolf2}
{\mathcal L \mathnormal} \left( \mbox{a} \right) &=& - g^2 w \sum_{\vec{\lambda}^{(l)}} \int_{{\rm{\bf p}}} \vec{\lambda}^{(l)2} \frac{1}{2 \kappa_{\vec{\lambda}+\vec{\lambda}^{(l)}} \kappa_{\vec{\lambda}}^2} 
\left[ \frac{1}{\kappa_{\vec{\lambda}} \left( \kappa_{\vec{\lambda}+\vec{\lambda}^{(l)}} + \kappa_{\vec{\lambda}}\right) } + \frac{1}{ \left( \kappa_{\vec{\lambda}+\vec{\lambda}^{(l)}} + \kappa_{\vec{\lambda}} \right)^2 } + {\sl O} \left( u,v\right) \right] \ .
\end{eqnarray}
In terms of the parameter sum, Eq.~(\ref{wolf2}) reads
\begin{eqnarray}
{\mathcal L \mathnormal} \left( \mbox{a} \right) = - g^2  w \frac{1}{2} \int_{{\rm{\bf p}}} \Big[ \Sigma \left( 1, 3, 1 \right) + \Sigma \left( 1, 2, 2 \right) + {\sl O} \left( n+m+k>5 \right) \Big] \ .
\end{eqnarray}
The terms with $n+m+k>5$ are convergent. We keep the divergent contributions and obtain 
\begin{eqnarray}
{\mathcal L \mathnormal} \left( \mbox{a} \right) = - \frac{1}{16} \, g^2 w \vec{\lambda}^{2} \int_{{\rm{\bf p}}} \frac{1}{\kappa_{\vec{0}}^5} \ .
\end{eqnarray}
The remaining momentum integration is straightforward. In real space one retrieves the result stated in Eq.~(\ref{realSpaceResult_a}).

The second diagram of the right-hand side of FIG.~\ref{specialDiagrams}, b,  is easier to compute. We mention only the result
\begin{eqnarray}
\label{a11}
\mbox{b} = - g^2 w \vec{\lambda}^2  \frac{2 G_\epsilon}{\epsilon} \tau^{-\epsilon /2} \delta_+ \left( z^\prime \right) \delta_+ \left( z^{\prime \prime} \right) \ .
\end{eqnarray}

\section{Evaluation of Diagrams for the ordinary transition}
\label{app:detailsOfCalOrdinary}
\begin{figure}[h]
\epsfxsize=2.3cm
\begin{eqnarray*}
\raisebox{-10mm}{\epsffile{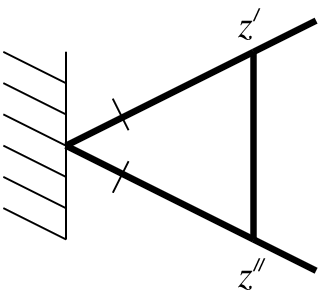}} = \ \raisebox{-11mm}{\epsffile{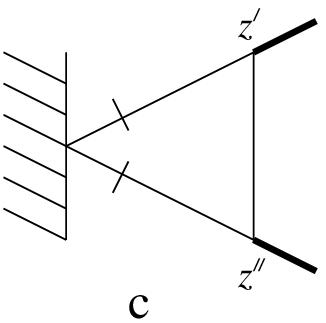}} - \ \raisebox{-11mm}{\epsffile{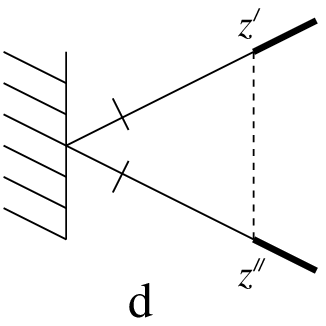}}
\end{eqnarray*}
\caption[]{\label{ordinaryDiagrams}Ultraviolet divergent Feynman diagrams contributing to the renormalization of $\mathcal{O}_{\mathcal S \mathnormal}^\infty$ at the ordinary transition. The meaning of the symbols is the same as in FIG.~\ref{specialDiagrams}, except that the hatched ''surface'' stands now, in conjunction with the bars, for an insertion of the operator ${\mathcal O \mathnormal}^{\infty}_{\mathcal S \mathnormal}$.}
\end{figure}

We start with diagram c displayed in FIG.~\ref{ordinaryDiagrams}. It stands for
\begin{eqnarray}
\label{b1}
\mbox{c} = -g^2 w \sum_{\vec{\lambda}^{(l)}} \int_{{\rm{\bf p}}} \vec{\lambda}^{(l)2} \left[ \partial_x G^{\mbox{\scriptsize{cond}},D}_{{\rm{\bf p}},\vec{\lambda}^{(l)}} \left( z^\prime, x \right) \right]_{x=0} \left[ \partial_y G^{\mbox{\scriptsize{cond}},D}_{{\rm{\bf p}},\vec{\lambda}^{(l)}} \left( y, z^{\prime \prime} \right) \right]_{y=0} G^{\mbox{\scriptsize{cond}},D}_{{\rm{\bf p}},\vec{\lambda} + \vec{\lambda}^{(l)}} \left( z^\prime, z^{\prime \prime} \right) \ ,
\end{eqnarray}
where the conducting Dirichlet propagator reads
\begin{eqnarray}
\label{b2}
G^{\mbox{\scriptsize{cond}},D}_{{\rm{\bf p}},\vec{\lambda}} \left( z, z^\prime \right) = \frac{1}{2 \kappa_{\vec{\lambda}} } \left[ \exp \left( - \kappa_{\vec{\lambda}} \left| z - z^\prime \right| \right) - \exp \left( - \kappa_{\vec{\lambda}} \left( z + z^\prime \right) \right) \right] \ .
\end{eqnarray}
Upon insertion of Eq.~(\ref{b2}) into Eq.~(\ref{b1}) one finds for the Laplace transformed of Eq.~(\ref{b1})
\begin{eqnarray}
\label{b3}
{\mathcal L \mathnormal} \left( \mbox{c} \right) &=& - g^2 w \sum_{\vec{\lambda}^{(l)}} \int_{{\rm{\bf p}}} \vec{\lambda}^{(l)2}
\frac{1}{2 \kappa_{\vec{\lambda}+\vec{\lambda}^{(l)}}}
\left[ \frac{1}{2\kappa_{\vec{\lambda}} + u + v} \left( \frac{1}{\kappa_{\vec{\lambda}+\vec{\lambda}^{(l)}} + \kappa_{\vec{\lambda}} + u} + \frac{1}{\kappa_{\vec{\lambda}+\vec{\lambda}^{(l)}} + \kappa_{\vec{\lambda}} + v}
\right)
\right.
\nonumber \\ 
&-& \left. \frac{1}{\kappa_{\vec{\lambda}+\vec{\lambda}^{(l)}} + \kappa_{\vec{\lambda}} + u} \, \frac{1}{\kappa_{\vec{\lambda}+\vec{\lambda}^{(l)}} + \kappa_{\vec{\lambda}} + v} \right] \ .
\end{eqnarray}
Next we carry out a Taylor expansion in terms of $1/\kappa_{\vec{\lambda}}$ and $1/ \left( \kappa_{\vec{\lambda}^{(l)}+\vec{\lambda}} + \kappa_{\vec{\lambda}} \right)$. We keep only those terms proportional to $uv$, since these are giving the leading behavior in the limit 
$c\to \infty$. They are in real space proportional to $\delta^\prime_+ \left( z^\prime \right) \delta^\prime_+ \left( z^{\prime \prime} \right)$, where the distribution $\delta^\prime_+ \left( z \right)$ is defined by
\begin{eqnarray}
\label{b4}
\int_0^\infty \delta^\prime_+ \left( z \right) g \left( z \right) = g^\prime \left( 0 \right) \ .
\end{eqnarray}
We obtain in terms of the parameter sum $\Sigma \left( n, m, k \right)$:
\begin{eqnarray}
\label{b5}
{\mathcal L \mathnormal} \left( \mbox{c} \right) = - g^2 w uv \int_{{\rm{\bf p}}} \left[ \frac{1}{4} \Sigma \left( 1, 3, 1 \right) + \frac{1}{4} \Sigma \left( 1, 2, 2 \right) - \frac{1}{2} \Sigma \left( 1, 0, 4 \right) + {\sl O} \left( n+m+k>5 \right) \right] \ .
\end{eqnarray}
Transforming back to real space yields
\begin{eqnarray} 
\label{b6}
\mbox{c} = - g^2 w \vec{\lambda}^2 \,  \frac{G_\epsilon}{10\epsilon} \,  \tau^{-\epsilon /2} \delta^\prime_+ \left( z^\prime \right) \delta^\prime_+ \left( z^{\prime \prime} \right) \ .
\end{eqnarray}
For the second diagram on the right-hand side of FIG.~\ref{ordinaryDiagrams} we obtain by similar means
\begin{eqnarray}
\label{b7}
\mbox{d} = - g^2 w \vec{\lambda}^2  \, \frac{5G_\epsilon}{6\epsilon} \, \tau^{-\epsilon /2} \delta^\prime_+ \left( z^\prime \right) \delta^\prime_+ \left( z^{\prime \prime} \right) \ .
\end{eqnarray}

\section{Evaluation of the parameter sum}
\label{app:evalParamSum}
Here we outline the evaluation of the parameter sum introduced in appendix~\ref{app:detailsOfCalSpecial}. We apply the inverse Mellin transformation\cite{erdelyi_54} 
\begin{eqnarray}
\label{m1}
\frac{1}{\left( \kappa_{\vec{\lambda}+\vec{\lambda}^{(l)}} + \kappa_{\vec{\lambda}} \right)^k} = \int_{\alpha -i\infty}^{\alpha -i\infty} \frac{d \sigma}{2\pi i} \frac{\Gamma \left( \sigma \right) \Gamma \left( k - \sigma \right) }{\Gamma \left( k \right)} \kappa_{\vec{\lambda}}^{\sigma -k} \kappa_{\vec{\lambda}+\vec{\lambda}^{(l)}}^{-\sigma} \ ,
\end{eqnarray}
for $0<\alpha <k$, to the right-hand side of Eq.~(\ref{parameterSum}). Using Schwinger parametrization we obtain
\begin{eqnarray}
\label{m2}
\Sigma \left( n, m, k \right) &=& \lim_{D\to 0} \int_{\alpha -i\infty}^{\alpha -i\infty} \frac{d \sigma}{2\pi i} \frac{\Gamma \left( \sigma \right) \Gamma \left( k - \sigma \right) }{\Gamma \left( \frac{n+\sigma}{2} \right) \Gamma \left( \frac{m+k-\sigma}{2} \right) \Gamma \left( k \right) } \int_0^\infty ds_1 \int_0^\infty ds_2 s_1^{(n+\sigma )/2 -1} s_2^{(m+k-\sigma )/2 -1} 
\nonumber \\
&\times&
\exp \left[ - \left( s_1 + s_2 \right) \left( \tau + {\rm{\bf p}} \right) \right] \left( - \frac{1}{w} \frac{\partial}{\partial s_2} \right) \sum_{\vec{\lambda}^{(l)}} \exp \left[ -  s_1 w \left( \vec{\lambda} + \vec{\lambda}^{(l)} \right)^2 - s_2 w \vec{\lambda}^{(l)2} \right]
\end{eqnarray}
Upon completion of the square the sum is easily carried out in the limit $D\to 0$. Moreover, we expand for small $\vec{\lambda}^2$. We obtain
\begin{eqnarray}
\label{m3}
\Sigma \left( n, m, k \right) &=& \vec{\lambda}^2 \int_{\alpha -i\infty}^{\alpha -i\infty} \frac{d \sigma}{2\pi i} \frac{\Gamma \left( \sigma \right) \Gamma \left( k - \sigma \right) }{\Gamma \left( \frac{n+\sigma}{2} \right) \Gamma \left( \frac{m+k-\sigma}{2} \right) \Gamma \left( k \right) } \int_0^\infty ds_1 \int_0^\infty ds_2 s_1^{(n+\sigma )/2 -1} s_2^{(m+k-\sigma )/2 -1} 
\nonumber \\
&\times&
\frac{s_1^2}{\left( s_1 + s_2 \right)^2}
\exp \left[ - \left( s_1 + s_2 \right) \left( \tau + {\rm{\bf p}} \right) \right] \ ,
\end{eqnarray}
up to terms of higher order in $\vec{\lambda}^2$. A change of variables $s_1 \to tx$, $s_2 \to t(1-x)$ renders the integration over the Schwinger parameters straightforward. One gets
\begin{eqnarray}
\label{m4}
\Sigma \left( n, m, k \right) = \vec{\lambda}^2 \frac{1}{\kappa_{\vec{0}}^{(n+m+l)}} \int_{\alpha -i\infty}^{\alpha -i\infty} \frac{d \sigma}{2\pi i}\frac{\Gamma \left( \sigma \right) \Gamma \left( k - \sigma \right) }{ \Gamma \left( k \right) } \frac{\left( n + \sigma \right) \left( n + \sigma +2 \right)}{\left( n + m + k  \right) \left( n + m + k + 2 \right)} \ .
\end{eqnarray}
The remaining integration can be done by exploiting the identity
\begin{eqnarray}
\label{m5}
\int_{\alpha -i\infty}^{\alpha -i\infty} \frac{d \sigma}{2\pi i}\frac{\Gamma \left( \sigma \right) \Gamma \left( k - \sigma \right) }{ \Gamma \left( k \right) } \sigma^{\nu} = \left( -t \frac{\partial}{\partial t}\right)^{\nu} \left. \int_{\alpha -i\infty}^{\alpha -i\infty} \frac{d \sigma}{2\pi i}\frac{\Gamma \left( \sigma \right) \Gamma \left( k - \sigma \right) }{ \Gamma \left( k \right) } t^{-\sigma} \right|_{t=1} \ .
\end{eqnarray}
Finally, one arrives at the result stated in Eq.~(\ref{resParameterSum}).


\newpage
%
%
\end{document}